\documentclass[11pt,a4paper]{article}

\usepackage{amssymb}
\usepackage{amsmath}
\usepackage{epsfig}
\usepackage{fancybox}
\usepackage{euler}
\usepackage[hang,stable]{footmisc}
\newcommand{\EM}[2]{\mathbf{EM}\{\,#1\mid #2\,\}}
\newcommand{\Lag}[2]{\mathbf{L}\,\{\,#1\mid #2\,\}}
\newcommand{\KPT}{\mathcal{K}}
\newcommand{\DPT}{\mathcal{D}}
\newcommand{\BAK}{\mathcal{B}}
\newcommand{\Gau}{G_{\mathrm{gau}}}
\newcommand{\Phy}{G_{\mathrm{obs}}}
\newcommand{\Parity}{\mathcal{P}}
\newcommand{\CC}{\mathcal{C}}
\renewcommand{\bar}{\overline}

\newtheorem{definition}{Definition}

\begin{document}
\ifx\href\undefined\else\hypersetup{linktocpage=true}\fi

\tolerance 10000

\title{Concepts of Symmetry\\ in the Work of Wolfgang Pauli}
\author{Domenico Giulini            \\
        Max-Planck-Institute for Gravitational Physics\\
        Albert-Einstein-Institute       \\
        Am M\"uhlenberg 1   \\
        D-14476 Golm, Germany}
\maketitle

\begin{abstract}
\noindent
``Symmetry'' was one of the most important methodological themes in 20th-century 
physics and is probably going to play no lesser role in physics of the 21st century. 
As used today, there are a variety of interpretations of this term, which differ in 
meaning as well as their mathematical consequences. Symmetries of crystals, for 
example, generally express a different kind of invariance than gauge symmetries, 
though in specific situations the distinctions may become quite subtle. I will 
review some of the various notions of ``symmetry'' and highlight some of 
their uses in specific examples taken from Pauli's scientific {\oe}vre. 

This paper is based on a talk given at the conference 
\emph{Wolfgang Pauli's Philosophical Ideas and Contemporary Science}, May 20.-25. 2007,
at Monte Verita, Ascona, Switzerland.
\end{abstract}
\newpage

\setcounter{tocdepth}{3}
\tableofcontents
\newpage

\section{General Introduction}
\label{sec:GenIntro}
In the Introduction to Pauli's Collected Scientific Papers, the editors, 
Ralph Kronig and Victor Weisskopf, make the following statement:
\begin{quote}
\emph{
It is always hard to look for a leading principle in the work of a 
great man, in particular if his work covers all fundamental problems 
of physics. Pauli's work has one common denominator: his striving
for symmetry and invariance. [...]
The tendency towards invariant formulations of physical laws, initiated
by Einstein, has become the style of theoretical physics in our days,
upheld and developed by Pauli during all his life by example, stimulation,
and criticism. For Pauli, the invariants in physics where the symbols 
of ultimate truth which must be attained by penetrating through the
accidental details of things. The search for symmetry and general validity
transcend the limits of physics in Pauli's work; it penetrated his thinking
and striving throughout all phases of his life, in all fields of philosophy 
and psychology.''}(\cite{Pauli:CSP}, Vol.\,1, p.\,viii)
\end{quote}
Indeed, if I were asked to list those of Pauli's scientific contributions
which make essential use of symmetry concepts and applied group theory, I 
would certainly include the following, which form a substantial 
part of Pauli's scientific {\oe}vre:\footnote{Two of the listed themes, 
``meson-nucleon interaction and differential geometry'' and 
``unifying non-linear spinor equation'', were never published in 
scientific journals (in the second case Heisenberg published for himself
without Pauli's consent) but can be followed from his letters and 
manuscripts as presented in \cite{Pauli:SC}.} 

\begin{quote}
Relativity theory and Weyl's extension thereof (1918-1921),
the Hydrogen atom in matrix mechanics (1925),
exclusion principle (1925),
anomalous Zeeman effect and electron spin (1925),
non-relativistic wave-equation for spinning electron (1927),
covariant QED (1928, Jordan),
neutrino hypothesis (1930),
Kaluza-Klein theory and its projective formulation (1933),
theory of $\gamma$-matrices (1935),
Poincar\'e-invariant wave equations (1939, Fierz),
general particle statistics and Lorentz invariance (1940, Belinfante),
spin-statistics (1940),
once more General Relativity and Kaluza-Klein theory (1943, Einstein),
meson-nucleon interaction and differential geometry (1953),
CPT theorem (1955),
$\beta$-decay and conservation of lepton charge (`Pauli group', 1957),
unifying non-linear spinor equation (collaboration with Heisenberg, 1957-58),
group structure of elementary particles (1958, Touschek).
\end{quote}

\begin{figure}
\begin{minipage}[c]{0.40\linewidth}
\epsfig{figure=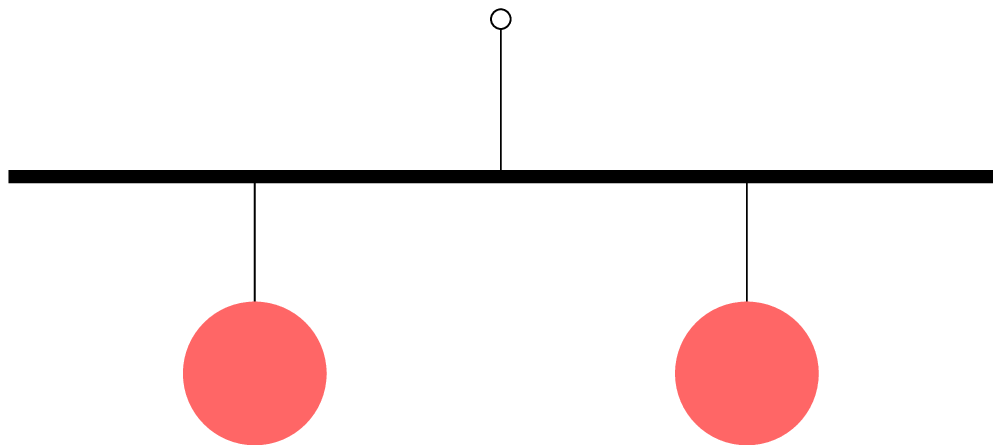, width=1.0\linewidth}
\put(-42.5,7){$M$}
\put(-112.5,7){$M$}
\put(-77,27){$m$}
\put(-75,68){$p$}
\end{minipage}
\hfill$\mathbf{\Longrightarrow}$\hfill
\begin{minipage}[c]{0.40\linewidth}
\epsfig{figure=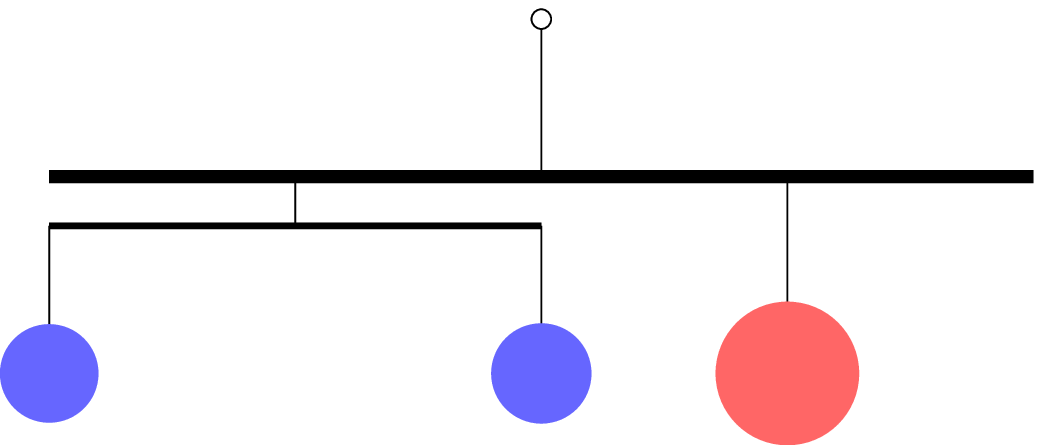, width=1.0\linewidth}
\end{minipage}\\[1cm]
\begin{minipage}[c]{0.40\linewidth}
\epsfig{figure=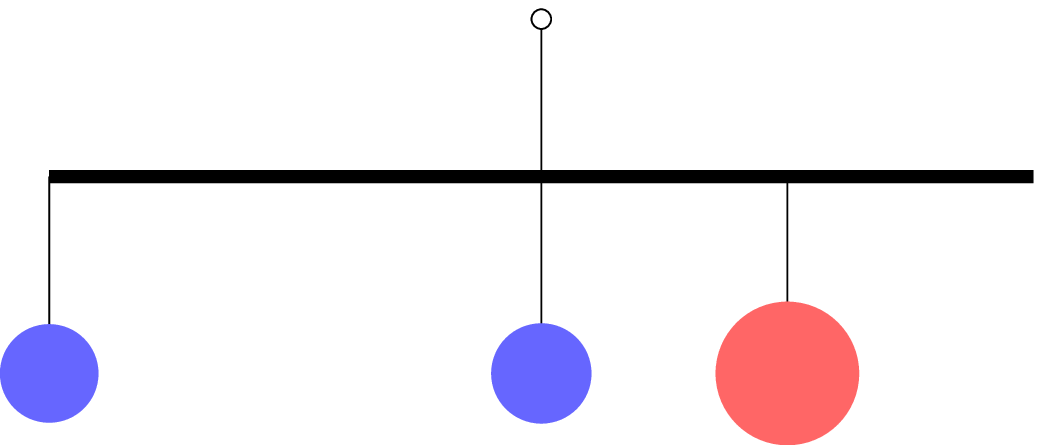, width=1.0\linewidth}
\end{minipage}
\hfill$\Longrightarrow$\hfill
\begin{minipage}[c]{0.40\linewidth}
\epsfig{figure=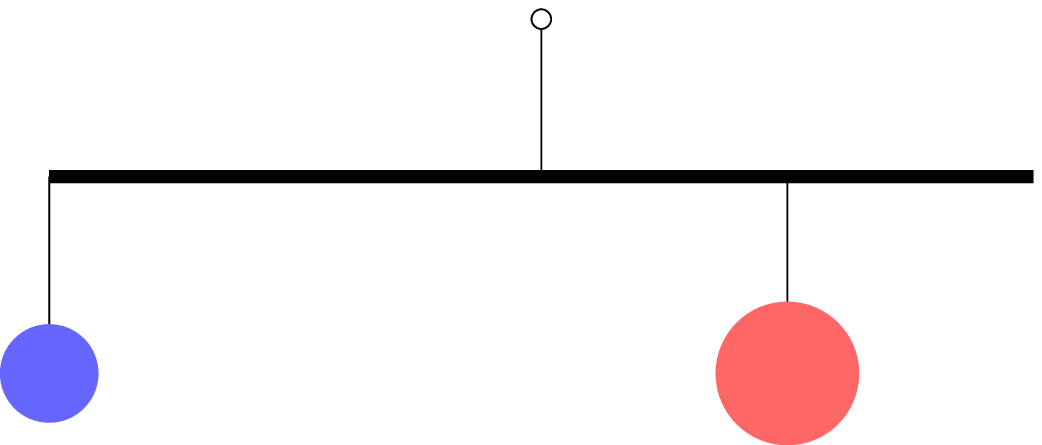, width=1.0\linewidth}
\end{minipage}
\caption{\label{fig:LawOfLever}%
The law of the lever `derived' from alleged symmetry considerations. 
The step from the upper right (second) to the lower left (third) picture
does not follow. The small (blue) balls represent half the mass of the 
big (red) balls.}
\end{figure}

Amongst the theoretical physicists of his generation, Pauli was certainly 
outstanding in his clear grasp of mathematical notions and methods. 
He had a particularly sober judgement of their powers as well as their 
limitations in applications to physics and other sciences. Let us once 
more cite Kronig and Weisskopf:
\begin{quote}
\emph{
Pauli's works are distinguished by their mathematical rigour and by a thorough and honest 
appraisal of the validity of assumptions and conclusions. He was a true disciple of Sommerfeld
in his clear mathematical craftsmanship. By example and sharp criticism he constantly tried to
maintain a similarly high standard in the work of other theoretical 
physicists. He was often called the living conscience of theoretical physicists.}
(\cite{Pauli:CSP}, Vol.\,1, p.\,viii)
\end{quote} 
It seems plausible that this critical impregnation dates back to his school-days,
when young Pauli read, for example, Ernst Mach's critical analysis of the 
historical development of the science of mechanics, a copy of which Pauli 
received as a present from his Godfather (Mach) at around the age of 
fourteen. Mach's ``Mechanik'', as this book is commonly called, starts out 
with a discussion of Archimedes' law of the lever, thereby criticising the 
following symmetry consideration (\cite{Mach:Mechanik}, p.\,11-12): 
Imagine two equal masses, $M$, and a perfectly stiff and homogeneous rod 
of length $L$, both being immersed into a static homogeneous vertical 
gravitational field, where the rod is suspended at its midpoint, $m$, 
from a point $p$ above; see Fig.\,\ref{fig:LawOfLever}. 
What happens if we attach the two equal masses to the ends of the rod and 
release them simultaneously without initial velocity? An immediate symmetry 
argument suggests that it stays horizontal; it might be given as follows: 
Everything just depends on the initial geometry and distribution of masses, 
which is preserved by a reflection at the plane perpendicular to the rod
through $p$ and $m$. Suppose that after release the rod dropped at one 
side of the suspension point $m$, then the mirror image of that process 
would have the same initial condition with the rod dropping to the other side. 
This is a contradiction if the laws governing the process are assumed to be 
reflection symmetric and deterministic (unique outcome for given initial 
condition). This argument seems rigorous and correct. Now, how does one get 
from here to the law of the lever? The argument criticised by Mach is as 
follows: Assume that the condition for equilibrium depends only on the amount 
of mass and its suspension point on the rod, but not on its shape. Then we 
may replace the mass to the left of $m$ by two masses of half the amount 
each on a small rod in equilibrium, as shown in the second (upper right) 
picture. Then replace the suspension of the small rod by two strings 
attached to the left arm of the original rod, as shown in the third 
(lower left) picture, and observe that the right one is just under the 
suspension point $m$, so that it does not disturb the equilibrium if it 
were cut away as in the last (lower right) picture of 
Fig.\,\ref{fig:LawOfLever}. 
The weak point in the argument is clearly the transition from the second to 
the third picture: There is no \emph{global} symmetry connecting them, even 
though 
locally, i.e. regarding the small rod only, it connects two equilibrium 
positions. It is easy to see that, in fact, the assumption that a global 
equilibrium is maintained in this change is \emph{equivalent} to 
Archimedes' law of the lever. This example shows (in admittedly a fairly 
trivial fashion) that alleged symmetry properties can work as a \emph{petitio
principii} for the law to be derived. This is essentially the criticism of 
Mach. 

The reason why we consider this `derivation' of the law of the lever to be 
a \emph{petitio principii} is that we have other, physically much more direct 
ways to actually derive it from \emph{dynamical} first principles. From that 
point of view the alleged symmetry is to be regarded as an artifact of the 
particular law and certainly not vice versa. The observed symmetry requires 
an explanation in terms of the dynamical laws, which themselves are to be 
established in an independent fashion. This is how we look upon, say, the 
symmetry of crystals or the symmetric shape of planetary orbits.  

On the other hand, all fundamental dynamical theories of 20th century 
physics are motivated by symmetry requirements. They are commonly looked
at as particularly simple realisations of the symmetries in question, given 
certain \emph{a priori} assumptions. It is clear that, compared to the previous
example, there are different concepts of symmetry invoked here. 
However, there also seems to be a shift in attitude towards a more 
abstract understanding of `physical laws' in general. 

What makes Pauli an interesting figure in this context is that 
this shift in attitude can be traced in his own writings. Consider 
Special Relativity as an example, thereby neglecting gravity. 
One may ask: What is the general relation between the particular 
symmetry (encoded by the Poincar\'e group) of spacetime and that 
very same symmetry of the fundamental interactions (weak, strong, 
and electromagnetic, but not gravity)? Is one to be considered as 
logically prior to the other? For example, if we take Einstein's 
original operationalist attitude, we would say that the geometry 
of spacetime is defined through the behaviour of `rods' and 
`clocks', which eventually should be thought of as physical systems 
obeying the fundamental dynamical laws. In fact, Einstein often 
complained about the fact that rods and clocks are introduced as 
if they were logically independent of the dynamical laws, e.g., 
in a discussion remark at the 86th meeting of the Gesellschaft 
Deutscher Naturforscher und \"Arzte in Bad Nauheim in 1920:%
\footnote{German original: ``Es ist eine logische Schw\"ache
der Relativit\"atstheorie in ihrem heutigen Zustande, da{\ss}
sie Ma{\ss}st\"abe und Uhren gesondert einf\"uhren mu{\ss},
statt sie also L\"osungen von Differentialgleichungen konstruieren 
zu k\"onnen.''}
\begin{quote}\emph{%
It is a logical shortcoming of the Theory of Relativity in its 
present form to be forced to introduce measuring rods and clocks 
separately instead of being able to construct them as solutions 
to differential equations.} 
(\cite{Einstein:CP}, Vol.\,7, Doc.\,46, p.\,353)
\end{quote}  
From that viewpoint, symmetry properties of spacetime are nothing 
but an effective codification of the symmetries of the 
fundamental laws. Consequences like `length contraction' and `time 
dilation' in Special Relativity are then only \emph{effectively} described 
as due to the geometry of spacetime, whereas a fundamental explanation 
clearly has to refer to the dynamical laws that govern clocks and rods. 
This was clearly the attitude taken by H.A.\,Lorentz and H.\,Poincar\'e,
though in their case still somehow afflicted with the idea of a 
material {\ae}ther that, in principle, defines a preferred rest frame, so 
that the apparent validity of the principle of relativity must be 
interpreted as due to a `dynamical conspiracy'.%
\footnote{H.A.\,Lorentz still expressed this viewpoint well after 
the formulation of Special Relativity, for example in 
\cite{Lorentz:DasRelativitaetsprinzip}, p.\,23.} In his famous 
article on Relativity for the Encyclopedia of Mathematical Sciences,
the young Pauli proposes to maintain this view, albeit without the idea 
on a material {\ae}ther. He writes:%
\footnote{German original: ``Ist aber das Bestreben, die 
Lorentz-Kontraktion atomistisch zu verstehen, vollkommen zu verwerfen?
Wir glauben diese Frage verneinen zu m\"ussen. Die Kontraktion des 
Ma{\ss}stabes ist kein elementarer, sondern ein sehr verwickelter Proze{\ss}.
Sie w\"urde nicht eintreten, wenn nicht schon die Grundgleichungen der 
Elektronentheorie sowie die uns noch unbekannten Gesetze, welche den 
Zusammenhalt des Elektrons selbst bestimmen, gegen\"uber der Lorentz-Gruppe
kovariant w\"aren. Wir m\"ussen eben postulieren, da{\ss} dies der Fall 
ist, wissen aber auch, da{\ss} dann, wenn dies zutrifft, die Theorie 
imstande sein wird, das Verhalten von bewegten Ma{\ss}st\"aben und Uhren 
atomistisch zu erkl\"aren.'' (\cite{Pauli:2000}, p.\,30.)}
\begin{quote}
\emph{Should one, then, in view of the above remarks, completely 
abandon any attempt to explain the Lorentz contraction 
atomistically? We think that the answer to this question 
should be No. The contraction of a measuring rod is not an 
elementary but a very complicated process. It would not take
place except for the covariance with respect to the Lorentz
group of the basic equations of electron theory, as well as 
those laws, as yet unknown to us, which determine the 
cohesion of the electron itself. We can only postulate that 
this is so, knowing that then the theory will be capable of 
explaining atomistically the behaviour of moving measuring 
rods and clocks.''} (\cite{Pauli:ToR-Dover}, p.\,15.)         
\end{quote}
Very recently, this traditional view has once more been defended 
under the name of `Physical Relativity'~\cite{Brown:PhysicalRelativity}
against todays more popular view, according to which Special Relativity 
is about the symmetry properties of spacetime itself. Clearly, the latter 
view only makes sense if spacetime is endowed with its own ontological 
status, independently of the presence of rods and clocks. 

This shift in emphasis towards a more abstract point of view is also 
reflected in Pauli's writings, for example in the Preface to the 
English edition of his `Theory of Relativity' of 1956, where the abstract 
group-theoretic properties of dynamical laws are given an autonomous status 
in the explanation of phenomena:   
\begin{quote}
\emph{
The concept of the state of motion of the `luminiferous {\ae}ther', 
as the hypothetical medium was called earlier, had to be given 
up, not only because it turned out to be unobservable, but 
because it became superfluous as an element of a mathematical 
formalism, the group-theoretical properties of which would only 
be disturbed by it. By the widening of the transformation group 
in general relativity the idea of a distinguished inertial 
coordinate system could also be eliminated by Einstein, being 
inconsistent with the group-theoretical properties of the theory.}
\end{quote}

Pushed to an extreme, this attitude results in the belief that the 
most fundamental laws of physics are nothing but realisations of 
basic symmetries. Usually this is further qualified by adding that 
these realisations are the most `simple' ones, at least with respect 
to some intuitive measure of simplicity. Such statements are 
well known from Einstein's later scientific period and also from 
Heisenberg in connection with his `unified theory' of elementary 
particles, for which he proposed a single non-linear differential 
equation, whose structure was almost entirely motivated by its 
symmetry properties. Heisenberg made this point quite explicitly 
in his talk entitled \emph{Planck's discovery
and the foundational issues of atomism}\footnote{German original: ``Die 
Plancksche Entdeckung und die philosophischen Grundfragen der Atomlehre''.},
delivered during the celebrations of Max Planck's 100th 
anniversary---at which occasion Wolfgang Pauli received the 
Max-Planck medal in absentia---, where he also talked about 
his own `unified theory':\footnote{German original:
``Die erw\"ahnte Gleichung enth\"alt neben den drei nat\"urlichen 
Ma{\ss}einheiten nur noch mathematische Symmetrieforderungen.
Durch diese Forderungen scheint alles weitere bestimmt zu sein. 
Man mu{\ss} eigentlich die Glei\-chung nur als eine besonders 
einfache Darstellung der Symmetrieforderungen, aber diese Forderung
als den eigentlichen Kern der Theorie betrachten.'' 
(\cite{Pauli:SC}, Vol.\,IV, Part\,IV\,B, p.\,1168)} 
\begin{quote}
\emph{%
The mentioned equation contains, next to the three natural 
units [$c,\hbar,l$], merely mathematical symmetry requirements. 
These requirements seem to determine everything else. In fact, 
one should just regard this equation as a particularly simple 
representation of the symmetry requirements, which form the 
actual core of the theory.}
\end{quote}
Pauli, who briefly collaborated with Heisenberg on this project, 
did not at all share Heisenberg's optimism that a consistent 
quantum-field theory could be based on Heisenberg's non-linear 
field equation. His objections concerned several serious technical 
aspects, overlayed with an increasing overall dislike of Heisenberg's 
readiness to make premature claims, particularly when made publicly. 

However, I think it is fair to say that the overall attitude regarding 
the heuristic r\^ole and power of symmetry principles in fundamental 
physics, expressed by Heisenberg in the above quote, was also to a large
extent shared by Pauli, not only in his later scientific life. This is 
particularly true for symmetry induced conservation laws, towards which 
Pauli had very strong feelings indeed. Examples from his later years 
will be discussed in later sections (e.g. Sect.\,\ref{sec:Irritations}). 
An example from his early scientific life is his strong resistance 
against giving up energy-momentum conservation for individual 
elementary processes, while keeping it on the statistical average.
Such ideas were advocated in the ``new radiation theory'' of Bohr, 
Kramers, and Slater of early 1924~\cite{Bohr.Kramers.Slater:1924} and  
again by Bohr in connection with $\beta$-decay, which Pauli called 
\emph{spiritual somersaults} in a letter to Max Delbr\"uck. 
A week after his famous letter suggesting the existence of the 
neutrino, Pauli wrote to Oskar Klein in a letter dated Dec. 12th 1930:%
\footnote{German original:
``Erstens scheint es mir, da{\ss} der Erhaltungssatz f\"ur Energie-Impuls
dem f\"ur die Ladung doch sehr weitgehend analog ist und ich kann keinen
theoretischen Grund daf\"ur sehen, warum letzterer noch gelten sollte (wie 
wir es ja empirisch f\"ur den $\beta$-Zerfall wissen), wenn ersterer 
versagt. Zweitens m\"u{\ss}te bei einer Verletzung des Energiesatzes auch 
mit dem \emph{Gewicht} etwas sehr merkw\"urdiges passieren. [...] 
Dies widerstrebt meinem physikalischen Gef\"uhl auf das \"au{\ss}erste!
Denn es mu{\ss} dann sogar auch f\"ur das Gravitationsfeld, das von 
dem ganzen Kasten (samt seinem radioaktiven Inhalt) selber \emph{erzeugt}
wird (...), angenommen werden, da{\ss} es sich \"andern kann, w\"ahrend
wegen der Erhaltung der Ladung das nach au{\ss}en erzeugte 
elektrostatische Feld (beide Felder scheinen mir doch analog zu sein; 
das wirst Du ja \"ubrigens auch aus deiner f\"unfdimensionalen 
Vergangenheit noch wissen) unver\"andert bleiben soll.'' 
(\cite{Pauli:SC}, Vol.\,II, Doc.\,[261], p.\,45-46)}
\begin{quote}
\emph{%
First it seems to me, that the conservation law for energy-momentum 
is largely analogous to that for electric charge, and I cannot see a 
theoretical reason why the latter should still be valid (as we know 
empirically from $\beta$-decay) if the former fails. Secondly, something 
strange should happen to the \emph{weight} if energy conservation fails. 
[...] This contradicts my physical intuition to an extreme! 
For then one has to even assume that the gravitational 
field \emph{produced} [...] by the box (including the 
radioactive content) can change, whereas the electrostatic 
field must remain unchanged due to charge conservation 
(both fields seem to me analogous; as you will remember 
from your five-dimensional past).}
\end{quote}
This is a truly remarkable statement. Not many physicists would 
nowadays dare suggesting such an intimate connection between the 
conservation laws of charge and energy-momentum. What Pauli hints 
at with his last remarks in brackets
is the Kaluza-Klein picture, in which electric charge is interpreted 
as momentum in an additional space dimension in a five-dimensional 
spacetime. 

It is not difficult to find explicit commitments from Pauli's later 
scientific life expressing his belief in the heuristic power of 
symmetry considerations. Let me just select two of them. The first 
is from his introduction to the International Congress of 
Philosophers, held in Z\"urich in 1954, where Pauli states:%
\footnote{German original: ``Es ist mir wahrscheinlich, dass die 
Tragweite des mathematischen Gruppenbegriffes in der Physik heute 
noch nicht ausgesch\"opft ist.'' (\cite{Pauli:CSP}, Vol.\,2, p.\,1345)} 
\begin{quote}
\emph{%
``It seems likely to me, that the reach of the mathematical group 
concept in physics is not yet fully exploited.''}
\end{quote}
The second is from his closing remarks as the president of the conference 
``50 Years of Relativity'' held in Berne in 1955, where with respect to 
the still unsolved problem of whether and how the gravitational field 
should be described in the framework of Quantum-Field-Theory he remarks:%
\footnote{German original: ``Es scheint mir also, da\ss\ nicht so sehr die 
Linearit\"at oder Nichtlinearit\"at der Kern der Sache ist, sondern eher 
der Umstand, da\ss\ hier eine allgemeinere Gruppe als die Lorentzgruppe
vorhanden ist.''(\cite{Pauli:CSP}, Vol.\,2, p.\,1306)} 
\begin{quote}
\emph{%
It seems to me, that the heart of the matter [the problem 
of quantising the gravitational field] is not so much the 
linearity or non-linearity, but rather the fact that there 
is present a more general group than the Lorentz group.}
\end{quote}
This, in fact, implicitly relates to much of the present-day research 
that is concerned with that difficult problem.   

Before we can discuss specific aspects of `symmetry' in Pauli's
work in Section\,\ref{sec:SpecificComments}, we wish to recall  
various aspects of symmetry principles as used in physics.

\section{Remarks on the notion of symmetry}
\label{sec:Symmetry} 

\subsection{Spacetime}
\label{sec:SymmetrySpacetime}
The term `symmetry' is used in such a variety of meanings, even in physics, 
that it seems appropriate to recall some of its its main aspects.
One aspect is that which mathematicians call an `automorphism' and which 
basically means a `structure preserving self-map'. Take as an example
(conceptually not an easy one) the modern notion of spacetime. First of 
all it is a set, $M$, the members of which are events, or better, `potential 
events', since we do not want to assume that every spacetime point to be 
an actual physical event in the sense that a material happening is taking
place, or at least not one which is dynamically relevant to the problem at 
hand.\footnote{Minkowski was well aware that empty domains of spacetime
may cause conceptual problems. Therefore, in his famous 1908 Cologne 
address \emph{Space and Time} (German original: ``Raum und Zeit''), 
he said: \emph{In order to not leave a yawning void, we wish to imagine that at every 
place and at every time something perceivable exists.} 
German original: ``Um nirgends eine g\"ahnende Leere zu lassen, wollen 
wir uns vorstellen, da{\ss} allerorten und zu jeder Zeit etwas Wahrnehmbares 
vorhanden ist''. (\cite{Minkowski:1909}, p.\,2)} That set is endowed 
with certain structures which are usually motivated through operational 
relations of actual physical events.

One such structure could be that of a preferred set of paths, which represent 
inertial (i.e. force free) motions of `test bodies', that is, localised 
objects which do not react back onto spacetime structure. This defines a 
so-called `path-structure' (compare \cite{Ehlers.Koehler:1977}%
\cite{Coleman.Korte:1980}), which in the simplest case reduces to an 
\emph{affine structure} in which the preferred paths behave, intuitively speaking, 
like `straight lines'. This can clearly be said in a much more precise 
form (see, e.g., \cite{Pfister:2004}). Under very mild technical assumptions 
(not even involving continuity) one may then show that the only automorphisms
of that `inertial structure' can already be narrowed down to the 
inhomogeneous Galilei or Lorentz groups, possibly supplemented by 
constant scale transformations 
(cf.~\cite{Giulini:2006b}\cite{Goldstein:2007}).\footnote{We shall from 
now on use `Poincar\'e group' for `inhomogeneous Lorentz group'
and `Lorentz group' for `homogeneous Lorentz group'.} 

Another structure to start with could have been that of a causal relation
on $M$. That is, a partial order relation which determines the pairs of 
points on spacetime which, in principle, could influence each other in form 
of a propagation process based on ordinary matter or light signals. 
The automorphism group of that structure is then the subgroup of bijections 
on $M$ that, together with their inverse, preserve this order relation. 
For example, in case of Minkowski space, where the causal relation is
determined by the light-cone structure, it may be shown that the most 
general automorphism is given by a Poincar\'e  transformation 
plus a constant rescaling\cite{Alexandrov:1975}\cite{Zeeman:1964}. 
Since, according to Klein's \emph{Erlanger Programm}~\cite{Klein:ErlangerProgramm}, 
any geometry may be characterised by its automorphism group, the 
geometry of Minkowski space is, up to constant rescalings, entirely 
encoded in the causal relations. 

The same result can be arrived at through topological considerations. 
Observers (idealised to be extensionless) move in spacetime on timelike 
curves. Take the set $\mathcal{C}$  of all (not necessarily smooth) 
timelike curves which 
are continuous in the standard (Euclidean) topology $\mathcal{T}_E$  
of Minkowski spacetime $M$. Now endow $M$ with a new topology, 
$\mathcal{T}_P$, called the \emph{path topology}, which is the finest 
topology on $M$ which induces the same topology on each path in 
$\mathcal{C}$ as the standard (Euclidean) topology $\mathcal{T}_E$. 
The new topology $\mathcal{T}_P$ is strictly finer than $\mathcal{T}_E$ 
and has the following remarkable property: The automorphism 
group of $(M,\mathcal{T}_P)$\footnote{In the standard topological way of 
speaking this is just the `homeomorphism group' of $(M,\mathcal{T}_P)$.}, 
i.e. the group of bijections of $M$ which, together with their inverses, 
preserve $\mathcal{T}_P$, is just the Poincar\'e group extended 
by the constant rescalings~\cite{Hawking.etal:1976}. This is possibly the 
closest operational meaning one could attribute to the topology of 
spacetime, since in $\mathcal{T}_P$ a set in spacetime is open if 
and only if every observer ``times'' it to be open. 

All this is meant to illustrate that there are apparently different 
ways to endow spacetime with structures that are, physically speaking, 
more or less well motivated and which lead to the same automorphism 
group. That group may then be called the group of \emph{spacetime symmetries}. 
So far, this group seems to bear no direct relation to any dynamical 
law. However, the physical meaning of such statements of symmetry is tight 
to an ontological status of spacetime points. We assumed from the 
onset that spacetime is a \emph{set} $M$. 
Now, recall that Georg Cantor, in his first article on transfinite set-theory 
\cite{CantorMengenlehre1:1895}, started out with the following 
definition of a set:\footnote{German original: ``Unter einer `Menge' 
verstehen wir jede  Zusammenfassung $M$ von bestimmten 
wohlunterschiedenen Objecten $m$  unserer Anschauung oder unseres 
Denkens (welche die `Elemente' von $M$ genannt werden) zu einem 
Ganzen.'' (\cite{CantorMengenlehre1:1895}, p.\,481)} 
\begin{quote}
\emph{
By a `set' we understand any gathering-together $M$ of determined well-distinguished 
objects $m$ of our intuition or of our thinking (which are called the `elements' of $M$)  
into a whole.}
\end{quote}
Hence we may ask: Is a point in spacetime, a `potential event' as we 
called it earlier, a ``determined well-distinguished object 
of our intuition or of our thinking''? This question is justified even
though modern axiomatic set theory is more restrictive in what may be 
called a set (for otherwise it runs into the infamous antinomies) and also 
stands back from any characterisation of elements in order to not confuse 
the axioms themselves with their possible \emph{interpretations}.\footnote{
This urge for a clean distinction between the axioms and their possible 
interpretations is contained in the famous and amusing dictum, attributed 
to David Hilbert by his student Otto Blumenthal: ``One must always be 
able to say 'tables', `chairs', and `beer mugs' instead of 'points, 
`lines', and `planes''. (German original: ``Man mu\ss\ jederzeit an Stelle 
von 'Punkten', `Geraden' und `Ebenen' 'Tische', `St\"uhle' und `Bierseidel' 
sagen k\"onnen.'')} However, applications to physics require interpreted 
axioms, where it remains true that elements of sets are thought of as 
definite as in Cantors original definition. But it is just this 
definiteness that seems to be physically unwarranted in application to 
spacetime. The modern general-relativistic viewpoint takes that into 
account by a quotient construction, admitting only those statements as 
physically meaningful that are invariant under the group of (differentiable) 
permutations of spacetime points. This is possible only because all other 
structures on spacetime, in particular the metric and with it the causal 
structure, are not fixed once and for all but are subsumed into the 
dynamical fields. Hence no non-dynamical background structures remain, 
except those that are inherent in the definition of a differentiable 
manifold. The group of automorphisms is therefore the whole diffeomorphism 
group of spacetime, which, in some sense, comes sufficiently close to the 
group of all permutations.\footnote{%
There are clearly much more general bijections of spacetime than continuous 
or even differentiable ones. However, the diffeomorphism group is still 
$n$-point transitive, that is, given any two $n$-tuples of mutually distinct
spacetime points, $(p_1,\cdots, p_n)$ and $(q_1,\cdots, q_n)$, there is a 
diffeomorphism $\phi$ such that for all $1\leq i\leq n$ we have 
$\Phi(p_i)=q_i$; this is true for all positive integers $n$.}

\subsection{Dynamical symmetries versus covariance}
\label{sec:DynamicalSymmetriesVersusCovariance}
What is the relation between spacetime automorphisms and symmetries of 
dynamical laws? Before we can answer this, we have to recall what a 
symmetry of a dynamical law is. 

For definiteness, let us restrict attention to dynamical laws in 
classical (i.e. non-quantum) physics.  
The equations of motion generally take the form of systems of differential 
equations, which we here abbreviate with $\mathbf{EM}$ 
($\mathbf{E}$quation of $\mathbf{M}$otion). These equations 
involve two types of quantities: 1) background structures, collectively 
abbreviated here by $\Sigma$, and 2) dynamical entities, collectively 
abbreviated here by $\Phi$. The former will typically be represented by geometric 
objects on $M$ (tensor fields, connections, etc), which are taken from 
a somehow specified set $\BAK$ of `admissible backgrounds'. Typical 
background structures are external sources, like currents, and the 
geometry of spacetime in non-general-relativistic field theories. 
Dynamical entities typically involve `particles' and `fields', which 
in the simplest cases are represented by maps to and from spacetime,
\begin{subequations}
\label{eq:ParticleField}
\begin{alignat}{4}
\label{eq:ParticleField-a}
&\gamma &&\,:\, \mathbb{R}&&\rightarrow M\qquad&&\text{(`particle')}\,,\\
\label{eq:ParticleField-b}
&\psi &&\,:\, M&&\rightarrow V\qquad &&\text{(`field')}\,,
\end{alignat}
\end{subequations}
and were $V$ is usually some vector space. 

In order to state the equations of motion, one has to first specify a set 
of so-called\footnote{This terminology is due to James Anderson~\cite{%
Anderson:RelativityPhysics}.}\textbf{kinematically possible trajectories} 
out of which the dynamical entities $\Phi$ are taken and solutions to the 
equations of motion are sought. Usually this involves particle 
trajectories which are sufficiently smooth (typically piecewise twice 
continuously differentiable) and fields which are sufficiently smooth
and in addition have a sufficiently rapid fall-off at large spatial 
distances, so as to give rise to finite quantities of energy, 
angular-momentum, etc. This space of kinematically possible 
trajectories will be denoted by $\KPT$. According to the discussion 
above, the equation of motion takes two arguments, one from $\BAK$ 
the other from $\KPT$, and is hence written in the form 
\begin{equation}
\label{eq:EOM}
\EM{\Sigma}{\Phi}=0\,,
\end{equation} 
where the zero on the right-hand side may be a many-component object. 
Equation (\ref{eq:EOM}) should be read as a selection criterion on 
the set $\KPT$, depending on the externally specified values of 
$\Sigma$. We shall sometimes write $\mathbf{EM}_\Sigma$ for 
$\EM{\Sigma}{\cdot}$ to denote the 
particular equation of motion for $\Phi$ corresponding to the 
choice $\Sigma$ for the background structures. In general, the sets 
of solutions to (\ref{eq:EOM}) for variable $\Sigma$ are $\Sigma$-dependent subset 
$\DPT_\Sigma\subset\KPT$, whose elements 
are called the 
\addtocounter{footnote}{-1}
\textbf{dynamically possible trajectories}\footnotemark. 
We can now say more precisely what is usually meant by a symmetry: 
\begin{definition}
\label{def:SymmetryGroup}
An abstract group $G$ is called a \textbf{symmetry group} of the
equations of motion iff\footnote{Throughout we use ``iff'' as 
abbreviation for ``if and only if''.} the following conditions
are satisfied:
\begin{enumerate}
\item
There is an effective (see below) action $G\times\KPT\rightarrow\KPT$ 
of $G$ on the set of kinematically possible trajectories, denoted by 
$(g,\Phi)\mapsto g\cdot\Phi$.
\item
This action leaves the subset $\DPT_\Sigma\subset\KPT$ invariant; that is,
for all $g$ in $G$ we have:
\begin{equation}
\label{eq:DefSymmetry}
\EM{\Sigma}{\Phi}=0\quad\Longleftrightarrow\quad \EM{\Sigma}{g\cdot\Phi}=0\,.
\end{equation}
\end{enumerate}
\end{definition} 
Recall that an action is called effective if no group element other than 
the group identity fixes all points of the set it acts on. Effectiveness is 
required in order to prevent mathematically trivial and physically 
meaningless extensions of $G$. What really matters are the orbits of $G$ in 
$\KPT$, that is, the subsets $O_\Phi=\{g\cdot\Phi\mid g\in G\}$ for each 
$\Phi\in\KPT$. If the action were not effective, we could simply reduce 
$G$ to a smaller group with an effective action and the same orbits in
$\KPT$, namely the quotient group $G/G'$, where $G'$ is the normal 
subgroup of elements that fix all points of $\KPT$. 

It should be noted that this definition is still very general due to 
the fact that no further condition is imposed on the action of $G$, 
apart from the obvious one of effectivity. For example, for fields  
one usually requires the action to be `local', in the sense 
that for any point $p$ of spacetime, the value $(g\cdot\psi)(p)$ of 
the $g$-transformed field should be determined by the value of the 
original field at some point $p'$ of spacetime, and possibly 
\emph{finitely many} derivatives of $\psi$ at $p'$. If there are no 
dependencies on the derivatives, the action is sometimes called 
`ultralocal'. Note that the point $p'$ need not be identical 
to $p$, but it is assumed to be uniquely determined by $g$ and $p$. 
A striking example of what can happen if locality is not imposed is 
given by the vacuum Maxwell equations (no external currents), which 
clearly admit the Poincar\'e group  
as ultralocally acting symmetry group. What is less well known is the 
fact that they also admit the inhomogeneous Galilei group as symmetry 
group\footnote{This is different from, and certainly more surprising 
than, the better known (ultra local) Galilei symmetry of Maxwell's 
equations in the presence of appropriate constitutive relations 
between the electric field $\vec E$ and the electric displacement-field 
$\vec D$ on one side,  and between the magnetic induction-field 
$\vec B$ and the magnetic field $\vec H$ on the other; see e.g.
\cite{LeBellac.LevyLeblond:1972} and \cite{Goldin.Shtelen:2001}.}, 
albeit the action is non-local; see \cite{Fushchich.Shtelen:1991} 
or Chap.\,5.9 of \cite{Fushchich:SymmDGL}. (There are also other non-local 
symmetries of the vacuum Maxwell equations~\cite{Fushchich.Nikitin:1979}.) 

To be strictly distinguished from the notion of symmetry is the notion 
of covariance, which we define as follows:
\begin{definition}
\label{def:CovarianceGroup}
An abstract group $G$ is called a \textbf{covariance group} of the
equations of motion iff the following conditions are satisfied:
\begin{enumerate}
\item
There is an effective action $G\times\KPT\rightarrow\KPT$ 
of $G$ on the set of kinematically possible trajectories, 
denoted by $(g,\Phi)\mapsto g\cdot\Phi$.
\item
There is also an action (this time not necessarily effective)
$G\times\BAK\rightarrow\BAK$ of $G$ on the set of background structures,  
likewise denoted by $(g,\Sigma)\mapsto g\cdot\Sigma$.
\item
The solution-function $\Sigma\mapsto\DPT_\Sigma\subset\KPT$ from $\BAK$
into the subsets of $\KPT$ is $G$-equivariant. This means the following: If 
$g\cdot\DPT_\Sigma$ denotes the set $\{g\cdot\Phi\mid\Phi\in\DPT_\Sigma\}$, 
then, for all $g$ in $G$, we have
\begin{equation}
\label{eq:DefCovariance1}
g\cdot\DPT_\Sigma=\DPT_{g\cdot\Sigma}\,.
\end{equation}
An alternative way to say this is that the relation that $\mathbf{EM}$ 
establishes on $\BAK\times\KPT$ via (\ref{eq:EOM}) is $G$ 
invariant, that is, for all $g$ in $G$, we have
\begin{equation}
\label{eq:DefCovariance2}
\EM{\Sigma}{\Phi}=0\quad\Longleftrightarrow\quad 
\EM{g\cdot\Sigma}{g\cdot\Phi}=0\,.
\end{equation}
\end{enumerate}
\end{definition}

The obvious difference between (\ref{eq:DefSymmetry}) and (\ref{eq:DefCovariance2})
is that in the former case the background structure is not allowed to change. 
The transformed dynamical entity is required to satisfy the \emph{very same}
equation as the untransformed one, whereas for a covariance it is only 
required to satisfy a suitably changed set of equations. Here `changed' 
refers to the fact that $g\cdot\Sigma$ is generally different from 
$\Sigma$. Hence it is clear that a symmetry group is automatically 
also a covariance group, by just letting it act trivially on the set 
$\BAK$ of background structures. The precise partial converse is as
follows: Given a covariance group $G$ with action on $\BAK$, then 
for each $\Sigma\in\BAK$ define the `stabiliser subgroup' of 
$\Sigma$ in $G$ as the set of elements in $G$ that fix $\Sigma$,
\begin{equation}
\label{eq:DefStab}
\mathrm{Stab}_G(\Sigma):=\{g\in G\mid g\cdot\Sigma=\Sigma\}\,.
\end{equation}
Then the subgroup $\mathrm{Stab}_G(\Sigma)$ of the covariance 
group is also a symmetry group of the equation of motion 
$\mathbf{EM}_\Sigma$. 

The requirement of covariance is a rather trivial one, since it can always 
be met by suitably taking into account all the background structures and a 
sufficiently general action of $G$ on $\BAK$. To see how this works in a 
specific example, consider the ordinary `heat equation' for the 
temperature field $T$ ($\kappa$ is a dimensionful constant):
\begin{equation}
\label{eq:HeatEq1}
\partial_t T-\kappa\Delta T=0\,.
\end{equation}
Let $G=E_3\times\mathbb{R}$ be the 7-parameter group of Euclidean motions
(rotations and translations in $\mathbb{R}^3$) and time translations, 
whose defining representation on spacetime ($\mathbb{R}^3\times\mathbb{R}$) 
is denoted by $g\rightarrow \rho_g$, then $G$ acts effectively on the 
set of temperature fields via $g\cdot T:=T\circ\rho_{g^{-1}}$ (the inverse 
being just introduced to make this a \emph{left} action). It is immediate from 
the structure of (\ref{eq:HeatEq1}) that this implements $G$ as symmetry 
group of this equation. The background structures implicit in  
(\ref{eq:HeatEq1}) are: a) a preferred split of spacetime into space and 
time, 2)~a preferred measure and orientation of time, and 
c)~a preferred distance measure on space. There are many ways to 
parametrise this structure, \emph{depending on the level of generality one starts from}. 
If, for example, we start from Special Relativity, we only list those 
structural elements that we need \emph{on top of} the Minkowski metric 
$\{\eta_{\mu\nu}\}=\text{diag}(1,-1,-1,-1)$ in order to write down 
(\ref{eq:HeatEq1}). They are given by a single constant and normalised 
timelike vector field $n$, by means of which we can write 
(\ref{eq:HeatEq1}) in the form 
\begin{equation}
\label{eq:HeatEq2}
\EM{n}{T}:=n^\mu\partial_\mu T
-\kappa(n^\mu n^\nu-\eta^{\mu\nu})\partial_\mu\partial_\nu T=0\,.
\end{equation}
In the special class of inertial reference frames in which 
$n^\mu=(1,0,0,0)$ equation (\ref{eq:HeatEq2}) reduces to (\ref{eq:HeatEq1}). 
From the structure of (\ref{eq:HeatEq2}) it is obvious that 
this equation admits the whole Poincar\'e group of Special 
Relativity as covariance group. However, the symmetry group it contains is 
the stabiliser subgroup of the given background structure. The latter 
is given by the vector field $n$, whose stabiliser subgroup within the 
Poincar\'e group is just $E_3\times\mathbb{R}$, the same as 
for (\ref{eq:HeatEq1}). 

Had we started from a higher level of generality, in which no preferred 
coordinate systems are given to us as in Special Relativity, we would 
write the heat equation in the form   
\begin{equation}
\label{eq:HeatEq3}
\EM{n,g}{T}:=n^\mu\nabla_\mu T
-\kappa(n^\mu n^\nu-g^{\mu\nu})\nabla_\mu\nabla_\nu T=0\,,
\end{equation}
where now $n$ as well as $g$ feature as background structures.  
$n$ is again specified as unit timelike covariant-constant vector 
field, $g$ as a flat metric, and $\nabla$ as the unique covariant 
derivative operator associated to $g$ (i.e. torsion free
and preserving $g$). Since $\nabla$ is here taken as a unique 
function of $g$, it does not count as independent background 
structure. Once again it is clear from the structure of (\ref{eq:HeatEq3}) 
that the covariance group is now the whole diffeomorphism group 
of spacetime. However, the symmetry group remains the same as 
before since the stabiliser subgroup of the pair $(g,n)$ is 
$E_3\times\mathbb{R}$. 

This example should make clear how easy it is to almost arbitrarily 
inflate covariance groups by starting from higher and higher levels 
of generality and adding the corresponding extra structures into ones 
list of background structures. This possibility is neither surprising 
nor particularly disturbing. Slightly more disturbing is the fact that
a similar game can be played with symmetries, at least on a very formal
level. The basic idea is to simply declare background structures to be  
dynamical ones by letting their values be determined by equations. 
We may do this since we have so far not qualified `equations of motion' 
as any special sort of equations. For example, in the special 
relativistic context we may just take (\ref{eq:HeatEq2}) and let $n$
be determined by 
\begin{equation}
\label{eq:HeatEq2-Suppl}
n^\mu n^\nu\eta_{\mu\nu}=1\,,
\qquad
\partial_\mu n^\nu=0\,.
\end{equation}
Then (\ref{eq:HeatEq2}) and (\ref{eq:HeatEq2-Suppl}) together define a 
background free (from the special relativistic point of view) system 
of equations for $T,n$ which has the full Poincar\'e group 
as symmetry group. Its symbolic form is 
\begin{equation}
\label{eq:HeatEq2-Symb}
\EM{\emptyset}{T,n}=0\,,
\end{equation}
where the $0$ on the right-hand side has now 18 components: one for 
(\ref{eq:HeatEq2}), one for the first equation in (\ref{eq:HeatEq2-Suppl}),
and 16 ($=4\times 4$) for the second equation in 
(\ref{eq:HeatEq2-Suppl}). But note that its $T$-sector of 
solution space is \emph{not} the same as that of (\ref{eq:HeatEq1}), 
as it now also contains solutions for different $n$.  However, as the 
equations (\ref{eq:HeatEq2-Suppl}) for $n$ do not involve $T$, the 
total solution space for $(n,T)$ can be thought of as fibred over 
the space of allowed $n$, with each fibre over $n$ being given by 
the solutions $T$ of (\ref{eq:HeatEq2}) for that given $n$. Each 
such fibre is a faithful image of the original solution space of 
(\ref{eq:HeatEq1}), suitably transformed by a Lorentz boost that 
relates the original $n$ in (\ref{eq:HeatEq1}) (i.e. 
$\{n^\mu\}=(1,0,0,0)$) to the chosen one. 

Even more radically, we could  take (\ref{eq:HeatEq3}) and declare $n$ 
and $g$ to be dynamical entities obeying the extra equations
\begin{equation}
\label{eq:HeatEq3-Suppl}
n^\mu n^\nu g_{\mu\nu}=1\,,
\qquad
\nabla_\mu n^\nu=0\,,
\qquad
\mathrm{Riem}[g]=0\,,
\end{equation}
where $\mathrm{Riem}$ is the Riemann curvature tensor of $g$, 
so that the last equation in (\ref{eq:HeatEq3-Suppl}) just 
expresses flatness of $g$.  The system consisting of (\ref{eq:HeatEq3}) 
and (\ref{eq:HeatEq3-Suppl}) has no background structures and admits 
the full diffeomorphism group as symmetry group. It is of the symbolic 
form  
\begin{equation}
\label{eq:HeatEq3-Symb}
\EM{\emptyset}{T,n,g}=0\,,
\end{equation}
which now comprises 36 components: the 16 as above and an additional set
of 20 for the independent components of $\mathrm{Riem}$.  Again, note that 
the $T$-sector of solution space of (\ref{eq:HeatEq3}) is now much 
bigger that of (\ref{eq:HeatEq1}) of or (\ref{eq:HeatEq2}). With 
any solution $T$ it also contains its diffeomorphism-transformed
one, $T'=T\circ\phi^{-1}$, where $\phi\in\mathrm{Diff}(M)$. Again, 
since the equations for $n$ and $g$ do not involve $T$, the total 
solution space is fibred over the allowed $n$ and $g$ fields, with 
each fibre corresponding to a faithful image of the original solution 
space for (\ref{eq:HeatEq1}).  
      
Finally we remark that, in principle, constants appearing in equations 
of motion could also be addressed as background structures whose values 
might eventually be determined by more general dynamical theories. 
For example, one might speculate (as was done some time ago in the 
so-called Brans-Dicke theories) that the gravitational constant is 
actually the value of some field that only in the present epoch of
our Universe has settled to a spatially constant and quasi-static 
value, but whose value at much earlier times was significantly different. 
Another example from Quantum Field Theory concerns the idea that 
masses of elementary particles are dynamically generated by the 
so-called Higgs field (whose existence is strongly believed but not 
yet experimentally confirmed). 

In any case, the important message from the considerations of 
this subsection is the following: symmetries emerge or disappear 
if, respectively, background structures become dynamical 
($\Sigma\rightarrow\Phi$) or dynamical structures `freeze' 
($\Phi\rightarrow\Sigma$).

\subsection{Observable versus gauge symmetries}
\label{sec:PhysicalVersusGaugeSymmetries}
Within the concept of symmetry as explained so far, an important 
distinction must be made between \emph{observable} symmetries on
one hand, and \emph{gauge} symmetries on the other.
An observable symmetry transforms a state or a history of states
(trajectory) into a \emph{different}, that is, \emph{physically 
distinguishable} state or history of states.  On the other hand, a 
gauge symmetry transforms a state or a history of states
into a \emph{physically indistinguishable} state or a history of states.
In this case there is a redundancy in the mathematical description,
so that the map from mathematical labels to physical states is not 
faithful. This is usually associated with a group, called the 
\emph{group of gauge transformations}, denoted by $\Gau$, which acts on 
the set of state labels such that two such labels correspond to 
the same physical state iff they lie in the same orbit of $\Gau$. 

It is clear that the notion of `distinguishability' introduced here 
refers to the set of observables, i.e. functions on state space
that are physically realisable in the widest sense. Assuming for the 
moment that this was well defined, we could attempt a definition 
as follows:
\begin{definition}
\label{def:ObservableSymmetry}
Let $G$ be a symmetry group in the sense of 
Definition\,\ref{def:SymmetryGroup}. Then $g\in G$ is called 
an \textbf{observable or physical symmetry} iff there exists 
a $\Phi\in\DPT_\Sigma$ and a physical observable that separates 
$g\cdot\Phi$ from $\Phi$. If no such observable exists, $g$ 
is called a \textbf{gauge symmetry}.       
\end{definition}
It is clear that for a theoretician the stipulation of what functions on
state space correspond to physically realisable observables is 
itself of hypothetical nature. However, what is important for us 
at this point is merely that \emph{relative} to such a stipulation the 
distinction between observables and gauge symmetries makes sense. 
In the mathematical practice gauge symmetries are often signalled by 
an underdeterminedness of the equations of motion, which sometimes 
simply fail to restrict the motion in certain degrees 
of freedom which are then called `gauge degrees of freedom'. In 
that case, given any solution $\Phi\in\DPT_\Sigma$, we can 
obtain another solution, $\Phi'$, by just changing $\Phi$ in those 
non-determined degrees of freedom in an arbitrary way. For example,
if the equations of motion are obtained via an action principle, 
such spurious degrees of freedom will typically reveal their nature 
through the property that motions in them are not associated 
with any action. As a result, the equations of motion, which are just 
the condition for the stationarity of the action, will not constrain 
the motion in these directions. Conversely, if according to the 
action principle the motion in some degree of freedom costs action, 
it can hardly be called a redundant one. In this sense an action 
principle is not merely a device for generating equations of motion,
but also contains some information about observables. 

The combination of observable and gauge symmetries into the total 
symmetry group $G$ need not at all be just that of a semi-direct 
or even direct product. Often, in field theory, the gauge group $\Gau$ 
is indeed a subgroup of $G$, in fact an invariant (normal) one, 
but the observable symmetries, $\Phy$, are merely a quotient and not 
a subgroup of $G$. In standard group theoretic terms one says that 
$G$ is a $\Gau-$extension of $\Phy$. This typically happens in 
electromagnetism or more generally in Yang-Mills type gauge theories 
or General Relativity with globally charged configurations. In this 
case only the `gauge transformations' with sufficiently rapid 
fall-off at large spatial distances are proper gauge transformations 
in our sense, whereas the long ranging ones cost action if performed 
in real time\footnote{By the very definition of global charge, which is 
just the derivative of the action with respect to a long-ranging 
'gauge transformation'.} and therefore have to be interpreted as 
elements of $\Phy$; see e.g. \cite{Giulini:1995d} and Chap.\,6 
of \cite{Joos-etal:2003}. 

This ends our small excursion into the realm of meanings 
of `symmetry'. We now turn to the discussion of specific 
aspects in Pauli's work.

\section{Specific comments on symmetries in Pauli's work}
\label{sec:SpecificComments}
The usage of symmetry concepts in Pauli's work is so rich and so 
diverse that it seems absolutely hopeless, and also inappropriate, 
to try to present them in a homogeneous fashion with any claim of 
completeness. Rather, I will comment on various subjectively selected 
aspects without in any way saying that other aspects are of any lesser 
significance. In fact, I will not include some of his most outstanding 
contributions, like, for example, the formulation of the exclusion 
principle, the neutrino hypothesis, or his anticipation of Yang-Mills 
Gauge Theory for the strong interaction. There exist excellent reviews and 
discussions of these topics in the literature. Specifically I 
wish to refer to Bartel van der Waerden's contribution 
\emph{Exclusion Principle and Spin} to the Pauli Memorial Volume 
(\cite{Fierz.Weisskopf:PauliMemorial}, pp.\,199-244), Norbert 
Straumann's recent lecture on the history of the exclusion 
principle~\cite{Straumann:ExclPrinc2004}, Pauli's own account of 
the history of the neutrino (in English: \cite{Pauli:WoPaP}, 
pp.\,193-217; in German: \cite{Pauli:CSP}, Vol.\,2, pp.\, 1313-1337 
and \cite{Pauli:PuE}, p.\,156-180), Chien-Shiung Wu's account 
\emph{The Neutrino} in the Pauli Memorial Volume 
(\cite{Fierz.Weisskopf:PauliMemorial}, pp.\,249-303),
and the historical account of gauge theories by  
Lochlainn O'Raifeartaigh and Norbert 
Straumann~\cite{Raifeartaigh.Straumann:2000}. A non-technical 
overview concerning \emph{Pauli's Belief in Exact Symmetries} is given 
by Karl von Meyenn~\cite{Meyenn:1987}. Last, but clearly not least, 
I wish to mention Charles Enz's fairly recent comprehensive 
scientific biography \cite{Enz:PauliBiography} of Wolfgang Pauli, 
which gives a detailed discussion of his scientific {\oe}vre.  

In this contribution I rather wish to concentrate on some particular
aspects of the notion of symmetry that are directly related to the 
foregoing discussion in Sections\,\ref{sec:DynamicalSymmetriesVersusCovariance} 
and \ref{sec:PhysicalVersusGaugeSymmetries}, as I feel that they 
are somewhat neglected in the standard discussions of symmetry.

\subsection{The hydrogen atom in matrix mechanics}
\label{sec:HydrogenAtom}
In January 1926 Pauli managed to deduce the energy spectrum for the 
Hydrogen atom from the rules of matrix mechanics. For this he implicitly 
used the fact that the mechanical problem of a point charge moving in 
a spherically symmetric force-field with a fall-off proportional to the 
square of the inverse distance has a symmetry group twice as large 
(i.e. of twice the dimension) as the group of spatial rotations alone, 
which it contains. Hence the total symmetry group is made half of 
a `kinematical' part, referring to space, and half of 
a `dynamical' part, referring to the specific force law ($1/r^2$ 
fall-off). Their combination is a proper physical symmetry group that 
transforms physically distinguishable states into each other. In the 
given quantum-mechanical context one also speaks of `spectrum generating' 
symmetries.     

Let us recall the classical problem in order to convey some idea where 
the symmetries and their associated conserved quantities show up, and 
how they may be employed to solve the 
dynamical problem. Consider a mass-point of mass $m$ and position coordinate 
$\vec r$ in the force field $\vec F(\vec r)=-(K/r^2)\vec n$, where $r$ 
is the length of $\vec r$, $\vec n:=\vec r/r$, and $K$ is some dimensionful
constant. Then, according to Newton's 3rd law (an overdot stands for the 
time derivative),     
\begin{equation}
\label{eq:EqMotSphSymPot}
\ddot{\vec r}=-\frac{k}{r^2}\vec n \qquad(k=K/m)\,.
\end{equation}

Next to energy, there are three obvious conserved quantities 
corresponding to the three components of the angular-momentum 
vector (here written per unit mass)
\begin{equation}
\label{eq:AngMomVect}
\vec\ell=\vec r\times\dot{\vec r}\,.
\end{equation}
But there are three more conserved quantities, corresponding to the 
components of the following vector (today called the Lenz-Runge vector),
\begin{equation}
\label{eq:LenzRungeVect}
\vec e=k^{-1}\dot{\vec r}\times\vec\ell-\vec n\,.
\end{equation}
Conservation can be easily verified by differentiation of 
(\ref{eq:LenzRungeVect}) using (\ref{eq:EqMotSphSymPot}) 
and $\dot{\vec n}=\vec\ell\times\vec n/r^2$. Hence on has 
($\ell$ = length of $\vec\ell$)
\begin{equation}
\label{eq:ConicSection1}
\vec\ell\cdot\vec r=0\,,\quad
\vec \ell\cdot\vec e=0\,,\quad
r+\vec r\cdot\vec e-k^{-1}\ell^2=0\,,
\end{equation}
from which the classical orbit immediately follows: Setting 
$\vec r\cdot\vec e=re\,\cos\varphi$, the last equation 
(\ref{eq:ConicSection1}) reads 
\begin{equation}
\label{eq:ConicSection2}
r=\frac{\ell^2/k}{1+e\,\cos\varphi}\,,
\end{equation}
which is the well known equation for a conic section in the plane 
perpendicular to $\vec\ell$, focus at the origin, eccentricity 
$e$ (= length of $\vec e$), and \emph{latus rectum} $2\ell^2/k$. 
The vector $\vec e$ points from the origin to the point of closest 
approach (periapsis). The few steps leading to this conclusion 
illustrate the power behind the method of working with conservation 
laws which, in turn, rests on an effective exploitment of symmetries. 

The total energy per unit mass is given by $E=\frac{1}{2}\dot{\vec r}^2-k/r$.
A simple calculation shows that  
\begin{equation}
\label{eq:ConicSection3}
e^2-1=2E\ell^2/k^2\,,
\end{equation}
which allows to express $E$ as function of the invariants 
$e^2$ and $\ell^2$.  This is the relation which Pauli shows 
to have an appropriate matrix analogue, where it allows to 
express the energy in terms of the eigenvalues of the matrices 
for $\ell^2$ and $e^2$ which Pauli determines, leading straight 
to the Balmer formula. 

From a modern point of view one would say that, for fixed energy 
$E<0$,\footnote{For $E>0$ one obtains a Hamiltonian action of 
$\mathfrak{so}(1{,}3)$.} the state space of this problem carries a 
Hamiltonian action of the Lie algebra $\mathfrak{so}(4)$, generated 
by the 3+3 quantities $\vec\ell$ and $\vec e$. Quantisation then 
consists in the problem to represent this Lie algebra as a commutator 
algebra of self-adjoint operators and the determination of spectra
of certain elements in the enveloping algebra. This is what Pauli
did, from a modern point of view, but clearly did not realise at 
the time. In particular, even though he calculated the commutation 
relations for the six quantities $\vec\ell$ and $\vec e$, he did not 
realise that they formed the Lie algebra for $\mathfrak{so}(4)$, as 
he frankly stated much later (1955) in his address on the occasion 
of Hermann Weyl's 70th birthday:\footnote{German original: 
``Ebensowenig wu{\ss}te ich, da{\ss} die Matrices, die ich ausgerechnet hatte,
um die Energiewerte des Wasserstoffatoms aus der neuen Quantenmechanik 
abzuleiten, eine Darstellung der 4-dimensionalen othogonalen Gruppe
gewesen sind''. (\cite{Pauli:SC}, Vol.\,IV, Part\,III, Doc.\,[2183], p.\,402)
Note that, in modern terminology, Pauli actually refers 
to a representation of the \emph{Lie algebra} of the orthogonal group.} 
\begin{quote}
\emph{%
Similarly I did not know that the matrices which I had derived from the 
new quantum mechanics in order to calculate the energy values of the 
hydrogen atom were a representation of the 4-dimensional orthogonal group.}
\end{quote}   
This may be seen as evidence for Pauli's superior instinct for detecting 
relevant mathematical structures in physics. Much later, in a
CERN-report of 1956, Pauli returned to the representation-theoretic 
side of this problem~\cite{Pauli:1965}.

\subsection{Particles as representations of spacetime automorphisms}
\label{sec:ParticlesRepresentations}
The first big impact of group theory proper on physics took place  
in quantum theory, notably through the work or Eugene 
Wigner~\cite{Wigner:GruppentheorieQM} and Hermann 
Weyl~\cite{Weyl:GrQM-1Ed}. While in atomic spectroscopy 
the usage of group theory could be looked upon merely as powerful 
mathematical tool, it definitely acquired a more fundamental flavour 
in (quantum) field theory. According to a dictum usually attributed to 
Wigner, every elementary system (particle) in special-relativistic 
quantum theory corresponds to a unitary irreducible representation of 
the Poincar\'e group.\footnote{The converse is not true, since there 
exist unitary irreducible representations which cannot correspond to 
(real) particles, for example the so-called `tachyonic' ones, 
corresponding to spacelike four-momenta.} In fact, all the 
Poincar\'e invariant linear wave equations on which special-relativistic 
quantum theory is based, known by the names of Klein \& Gordon, Weyl, 
Dirac, Maxwell, Proca, Rarita \& Schwinger, Bargmann \& Wigner, 
Pauli \& Fierz, can be understood as projection conditions that 
isolate an irreducible sub-representation of the Poincar\'e
group\footnote{%
More precisely, its 
universal cover $\mathbb{R}^4\rtimes\mathrm{SL}(2{,}\mathbb{C})$, 
or sometimes an extension thereof by the discrete transformations of 
space and time reversal.} within a reducible one that is easy to 
write down. More concretely, the latter is usually obtained as follows: 
Take a field $\psi$ on spacetime $M$ with values in a finite-dimensional 
complex vector space $V$. Let $D$ be a finite-dimensional irreducible 
representation of the (double cover of the) Lorentz 
group $\mathrm{SL}(2{,}\mathbb{C})$ on $V$.\footnote{The representation 
$D$ is never unitary, simply because the Lorentz group has no 
non-trivial finite-dimensional unitary irreducible representations. 
But it will give rise to an infinite-dimensional representation on the 
linear space of fields $\psi$ which will indeed be unitary.} It is 
uniquely labelled by a pair $(p,q)$ of two positive integer- or 
half-integer-valued numbers. In the standard terminology, $2p$ 
corresponds to the number of unprimed, $2q$ to the number of 
primed spinor indices of $\psi$. The set of such fields furnishes 
a linear representation of the (double cover of the) Poincar\'e group, 
$\mathbb{R}^4\rtimes\mathrm{SL}(2{,}\mathbb{C})$, where the action 
of the group element $g=(a,A)$ is given by 
\begin{equation}
\label{eq:ActionOnField}
g\cdot\psi:=D(A)(\psi\circ g^{-1})\,,
\end{equation} 
or for the Fourier transform $\tilde\psi$,
\begin{equation}
\label{eq:ActionOnField-FT}
g\cdot\tilde\psi:=\exp(ip_\mu a^\mu) D(A)(\tilde\psi\circ A^{-1})\,.
\end{equation} 

One immediately infers from (\ref{eq:ActionOnField-FT}) 
that irreducibility implies that $\tilde\psi$ must have support 
on a single $\mathrm{SL}(2{,}\mathbb{C})$ orbit in momentum space.
Here one usually restricts to those orbits consisting of non-spacelike 
$p$ (those with spacelike $p$ give rise to the tachyonic representations
which are deemed unphysical), which are labelled by $p_\mu p^\mu=m^2$ 
with non-negative $m$. For $\psi$ this means that it obeys the 
Klein-Gordon equation $(\Box+m^2)\psi=0$. This is already half the 
way to an irreducible representation, insofar as it now contains only 
modes of fixed mass. But these modes still contains several spins up 
to the maximal value $p+q$. A second and last step then consists of 
projecting out one (usually the highest) spin, which gives rise to 
the equations named above. In this fashion the physical meanings 
of \emph{mass} 
and \emph{spin} merge with the abstract mathematical meaning of 
mere labels of irreducible representations. Mass and spin are the 
most elementary attributes of physical objects, so that objects 
with no other attributes are therefore considered \emph{elementary}.
As just described, these elementary attributes derive from 
the representation theory of a group whose significance is usually 
taken to be that it is the automorphism group of spacetime. However, 
as already discussed in Sections\,\ref{sec:GenIntro} and 
\ref{sec:SymmetrySpacetime}, this point of view presupposes a 
hierarchy of physical thinking in which spacetime (here Minkowski 
space) is considered an entity prior to (i.e. more fundamental than) 
matter, which may well be challenged. A more consistent but 
also more abstract point of view would be to think of the  
abstract\footnote{`Abstract' here means to consider the 
isomorphicity class of the group as mathematical structure, 
without any interpretation in terms of transformations of an 
underlying set of objects.} Poincar\'e group as prior to the matter 
content \emph{as well as} the  spacetime structure and to derive 
both simultaneously. Here `deriving' a spacetime structure 
(geometry) from a group would be meant in the sense of Klein's 
\emph{Erlanger Programm}~\cite{Klein:ErlangerProgramm}.

We have already discussed in Section\,\ref{sec:GenIntro} Pauli's 
shift in emphasis towards a more abstract point of view as regards 
spacetime structure. But also as regards to matter he was, next 
to Wigner, one of the proponents to put symmetry considerations 
first and to derive the wave equations of fundamental fields as 
outlined above. Based on previous work by Fierz on the theory of 
free wave equations for higher spin~\cite{Fierz:1939}, Fierz and 
Pauli published their very influential paper \emph{On Relativistic Equations
for Particles of Arbitrary Spin in an Electromagnetic Field} 
(\cite{Pauli:CSP}, Vol.\,2, pp.\,873-894) which is still much 
cited today. 

In fact, much earlier, in his 1927 paper 
\emph{Quantum Mechanics of the Magnetic Electron}\footnote{German original:
``Zur Quantenmechanik des magnetischen Elektrons''. 
(\cite{Pauli:CSP}, Vol.\,2, pp.\,306-330)}, Pauli succeeded to implement 
the electron's spin into non-special-relativistic quantum mechanics 
in an entirely representation-theoretic fashion as regards the (Lie 
algebra of) spatial rotations. In contrast to the other (translational) 
degrees or freedom, spin does not appear as the quantisation of an already 
existent classical degree of freedom. This must have appeared 
particularly appealing to Pauli, who never wanted the electron's
`spin' to be understood as an intrinsic angular momentum due to 
a spatial rotation of a material structure. When Pauli introduced 
the new spin quantum-number for the electron in his 1924 paper 
\emph{On the Influence of the Velocity Dependence of the Electron Mass on
the Zeeman Effect}\footnote{German original: ``\"Uber den Einflu{\ss}
der Geschwindigkeitsabh\"angigkeit der Elektronenmasse auf den Zeemaneffekt''.
(\cite{Pauli:CSP}, Vol.\,2, pp.\,201-213)} he deliberately stayed 
away from any model interpretation and cautiously referred to it as 
\emph{a peculiar, classically indescribable disposition of two-valuedness of the 
quantum-theoretic properties of the light-electron}\footnote{German original:
``eine eigent\"umliche, klassisch nicht beschreibbare Art von 
Zweideutigkeit der quantentheoretischen Eigenschaften des 
Leuchtelektrons'' (\cite{Pauli:CSP}, Vol.\,2, p.\,213).}.
At that time an understandable general scepticism against possible 
erroneous prejudices imposed by the usage of classical models had 
already firmly established itself in Pauli's (and others) thinking. 

As much justified as this is in view of Quantum Mechanics, this 
had also led to overstatements to the effect that spin has no 
classical counterpart and that any classical model is even 
classically contradictory in the sense of violating Special Relativity. 
As regards the second point, which was also pushed by Pauli, we 
refer to the detailed discussion in \cite{Giulini:2007c}. 
To the first point we first wish to mention that composite models 
with half-integer angular momentum states exist in ordinary 
Quantum Mechanics (without spin), as, e.g.,  pointed out by Bopp \& Haag 
in 1950 \cite{Bopp.Haag:1950}. This is possible if their classical  
configuration space contains the whole group $SO(3)$ of spatial 
rotations. Pauli himself showed in his 1939 paper \emph{On a Criterion for Single- or 
Double-Valuedness of the Eigenfunctions in Wave Mechanics}\footnote{German original:
``\"Uber ein Kriterium f\"ur Ein- oder Zweiwertigkeit der Eigenfunktionen 
in der Wellenmechanik''. (\cite{Pauli:CSP}, Vol.\,2, pp.\,847-868)}
the possibility of double-valued wavefunctions, which are the ones 
that give rise to half-integer angular momentum states. 
Moreover, in classical mechanics there is also a precise analog 
of Wigner's notion of an elementary system. Recall that the space 
of states of a mechanical system is a symplectic manifold (phase space). 
The analog of an irreducible and unitary representation of the group 
of spacetime automorphisms is now a transitive and Hamiltonian action 
of this group on the symplectic manifold. It is interesting to note 
that this classical notion of an elementary system was only formulated 
much later than, and in the closest possible analogy with, the quantum 
mechanical one. An early reference where this is spelled out is 
\cite{Bacry:1967}. The classification of elementary systems is now 
equivalent to the classification of symplectic manifolds admitting such 
an action. An early reference where this has been done is~\cite{Arens:1971a}.
Here, as expected, an intrinsic angular momentum shows up as 
naturally as it does in Quantum Mechanics. What makes it slightly 
unusual (but by no means awkward or even inconsistent) is the fact 
that it corresponds to a phase space\footnote{The phase space for 
classical spin is a 2-sphere, which is compact and therefore leads 
to a finite-dimensional Hilbert space upon quantisation.} that is not 
the cotangent bundle (space of momenta) over some configuration 
space of positions.

Pauli's later writings also show this strong inclination to 
set the fundamentals of (quantum) field theory in 
group-theoretic terms. In his survey \emph{Relativistic Field Theories 
of Elementary Particles} (\cite{Pauli:CSP}, Vol.\,2, pp.\,923-952), 
written for the 1939 Solvay Congress, Pauli immediately starts a 
discussion of  ``transformation properties of the field equations 
and conservation laws''. His posthumously published notes on 
\emph{Continuous Groups in Quantum Mechanics}~\cite{Pauli:1965} focus 
exclusively on Lie-algebra methods in representation theory.

Today we are used to \emph{define} physical quantities like energy, 
momentum, and angular momentum as the conserved quantities associated 
to spacetime automorphisms via Noether's theorem. Here, too, Pauli 
was definitely an early advocate of this way of thinking. Reviews on 
the subject written shortly after Pauli's death show clear traces 
of Pauli's approach; see e.g.~\cite{Kemmer.etal:1959}.

\subsection{Spin and statistics}
\label{sec:SpinStatistics}
Pauli's proof of the spin-statistics correlation \cite{Pauli:1940}
(also \cite{Pauli:CSP}, Vol.\,2, pp.\,911-922), first shown by 
Markus Fierz in his habilitation thesis \cite{Fierz:1939}, is a truly 
impressive example for the force of abstract symmetry principles.
Here we wish to recall the basic lemmas on which it rests, which 
merely have to do with classical fields and representation theory.  

We begin by replacing the proper orthochronous Lorentz group 
by its double (= universal) cover $\mathrm{SL}(2{,}\mathbb{C})$ 
in order to include half-integer spin fields. We stress that 
everything that follows merely requires the invariance under 
this group. No requirements concerning invariance under 
space- or time reversal are needed!

We recall from the previous section that any finite-dimensional 
complex representation of $\mathrm{SL}(2{,}\mathbb{C})$ is labelled 
by an ordered pair $(p,q)$, where $p$ and $q$ may assume independently 
all non-negative integer or half-integer values. $2p$ and $2q$ correspond 
to the numbers of `unprimed' and `primed' spinor indices, respectively.
The tensor product of two such representations decomposes as follows:
\begin{equation}
\label{eq:ClebschGordan}
D^{(p,q)}\otimes D^{(p',q')}=
\bigoplus_{r=\vert p-p'\vert}^{p+p'}\quad
\bigoplus_{s=\vert q-q'\vert}^{q+q'}\, D^{(r,s)}\,,
\end{equation}
where---and this is the important point in what follows---the sums
proceed in \emph{integer} steps in $r$ and $s$.
With each $D^{(p,q)}$ let us associate a `Pauli Index',
given by
\begin{equation}
\label{eq:PauliIndex}
\pi: D^{(p,q)}\rightarrow ((-1)^{2p}\,,\,(-1)^{2q})
\ \in\ \mathbb{Z}_2\times\mathbb{Z}_2\,.
\end{equation}
This association may be extended to sums of such $D^{(p,q)}$
proceeding in integer steps, simply by assigning to the sum
the Pauli Index of its terms (which are all the same).
Then we have\footnote{This may be expressed by saying that the
map $\pi$ is a homomorphism of semigroups. One semigroup consists of
direct sums of irreducible representations proceeding in integer
steps with operation $\otimes$, the other is
$\mathbb{Z}_2\times\mathbb{Z}_2$, which is actually a group.}
\begin{equation}
\label{eq:PauliIndexTensor}
\pi(D^{(p,q)}\otimes D^{(p',q')})=\pi(D^{(p,q)})\cdot\pi(D^{(p',q')})\,.
\end{equation}

According to their representations, we can associate a Pauli Index with
spinors and tensors. For example, a tensor of odd/even degree
has Pauli Index $(-,-)$/$(+,+)$. The partial derivative, $\partial$,
counts as a tensor of degree one. Now consider the most general linear
(non interacting) field equations for integer spin (here and in what
follows $\sum (\cdots)$ simply stands for ``sum of terms of the
general form $(\cdots)$''):
\begin{equation}
\label{eq:FreeFieldEq}
\begin{split}
\sum\partial_{(-,-)}\Psi_{(+,+)}&\,=\,\sum \Psi_{(-,-)}\,,\\
\sum\partial_{(-,-)}\Psi_{(-,-)}&\,=\,\sum \Psi_{(+,+)}\,.\\
\end{split}
\end{equation}
These are invariant under
\begin{equation}
\label{eq:Theta}
\Theta:
\begin{cases}
\Psi_{(+,+)}(x)\ \mapsto\ +&\hspace{-2mm}\Psi_{(+,+)}(-x),\\
\Psi_{(-,-)}(x)\ \mapsto\ -&\hspace{-2mm}\Psi_{(-,-)}(-x)\,.
\end{cases}
\end{equation}
Next consider any current that is a polynomial in the fields
and their derivatives:
\begin{equation}
\label{eq:Current}
\begin{split}
J_{(-,-)}= \sum\ &\Psi_{(-,-)}
 + \Psi_{(+,+)}\Psi_{(-,-)}
 + \partial_{(-,-)}\Psi_{(+,+)} \\
 + &\Psi_{(+,+)}\partial_{(-,-)}\Psi_{(+,+)}
 + \Psi_{(-,-)}\partial_{(-,-)}\Psi_{(-,-)}
 + \quad\cdots\\
\end{split}
\end{equation}
Then one has
\begin{equation}
(\Theta J)(x)=-J(-x)\,.
\end{equation}
This shows that for any solution of the field equations with charge $Q$
for the conserved current $J$ ($Q$ being the space integral over $J^0$) there
is another solution (the $\Theta$ transformed) with charge $-Q$.
It follows that charges of conserved currents cannot be sign-definite in
any $\mathrm{SL}(2{,}\mathbb{C})$-invariant theory of non-interacting
integer spin fields. In the same fashion one shows that
conserved quantities, stemming from divergenceless symmetric
tensors of rank two, bilinear in fields, cannot be sign-definite
in any $\mathrm{SL}(2,\mathbb{C})$ invariant theory of non-interacting
half-integer spin fields. In particular, the conserved quantity in
question could be energy!

An immediate but far reaching first conclusion (not explicitly drawn 
by Pauli) is that there cannot exist a relativistic generalisation 
of Schr\"odinger's one- particle wave equation. For example, for 
integer-spin particles, one simply cannot construct a non-negative 
spatial probability distribution derived from conserved four-currents. 
This provides a general argument for the need of second 
quantisation, which in textbooks is usually restricted to the 
spin-zero case.  

Upon second quantisation the celebrated spin-statistics connection for 
free fields can now be derived in a few lines. It says that integer 
spin fields cannot be quantised using anti-commutators and half-integer 
spin field cannot be quantised using commutators. Here the so-called 
Jordan-Pauli distribution plays a crucial role\footnote{The Jordan-Pauli 
distribution was introduced by Jordan and Pauli in their 1927 paper
\emph{Quantum Electrodynamics of Uncharged Fields} (``Zur Quantenelektrodynamik 
ladungsfreier Felder''; \cite{Pauli:CSP}, Vol.\,2, pp.\,331-353) in 
an attempt to formulate Quantum Electrodynamics in a manifest Poincar\'e 
invariant fashion. It is uniquely characterised 
(up to a constant factor) by the following requirements: (1) it must be 
Poincar\'e invariant under simultaneous transformations of both arguments; 
(2) it vanishes for spacelike separated arguments;    
(3) it satisfies the Klein-Gordon equation. The (anti)commutators of 
the free fields must be proportional to the Jordan-Pauli distribution, 
or to finitely many derivatives of it.} in the (anti)commutation 
relations, which ensures causality (observables localised in spacelike 
separated regions commute). Also, the crucial hypothesis of the existence 
of an $\mathrm{SL}(2,\mathbb{C})$ invariant stable vacuum state is 
adopted. Pauli ends his paper by saying:
\begin{quote}
\emph{In conclusion we wish to state, that according to our opinion
the connection between spin and statistics is one of the most
important applications of the special relativity theory.}
(\cite{Pauli:1940}, p.\,722)
\end{quote}
It took almost 20 years before first attempts were made to 
generalise this result to the physically relevant case of interacting 
fields by L\"uders \& Zumino~\cite{Lueders.Zumino:1958}.

\subsection{The meaning of `general covariance'}
\label{sec:GeneralCovariance}
General covariance is usually presented as \emph{the} characteristic 
feature of General Relativity. The attempted meaning is that a generally 
covariant law takes the `same form' in all spacetime coordinate systems.
However, in order to define the `form' of a law one needs to make precisely 
the distinction between background entities, which are constitutive 
elements of the law, and the dynamical quantities which are to be obey 
the laws so defined (cf. Section\,\ref{sec:SymmetrySpacetime}). 
In the language we introduced above, `general covariance' cannot just 
mean simple covariance under all smooth and invertible transformations 
of spacetime points, i.e. that the spacetime diffeomorphism 
group is a covariance group is the sense of 
Definition\,\ref{def:CovarianceGroup}, for that would be easily achievable 
without putting any restriction on the intended law proper, as was already 
pointed out by Erich Kretschmann in 1917~\cite{Kretschmann:1917}. Einstein 
agreed with that criticism of Kretschmann's, which he called ``acute'' 
(German original: ``scharfsinnig'')(\cite{Einstein:CP}, Vol.\,7, Doc.\,4, 
pp.\,38-41), and withdrew to the view that the principle of general 
covariance has at least some heuristic power in the following sense:%
\footnote{German original: ``Von zwei mit der Erfahrung vereinbaren 
theoretischen Systemen wird dasjenige zu bevorzugen sein, welches
vom Standpunkte des absoluten Differentialkalk\"uls das einfachere
und durchsichtigere ist. Man bringe einmal die Newtonsche 
Gravitationsmechanik in die Form von kovarianten Gleichungen 
(vierdimensional) und man wird sicherlich \"uberzeugt sein, da\ss\
das Prinzip a) diese Theorie zwar nicht theoretisch, aber praktisch 
ausschlie{\ss}t.'' (\cite{Einstein:CP}, Vol.\,7, Doc.\,4, 
p.\,39)}
\begin{quote}
\emph{
Between two theoretical systems which are compatible with experience, 
that one is to be preferred which is the simpler and more transparent 
one from the standpoint of the absolute differential calculus. 
Try to bring Newton's gravitational mechanics in the form of generally 
covariant equations (four dimensional) and one will surely be convinced 
that principle a)\footnote{Einstein formulates principle a) thus:
``Principle of relativity: The laws of nature exclusively contain 
statements about spacetime coincidences; therefore they find their 
natural expression in generally covariant equations.'' 
(\cite{Einstein:CP}, Vol.\,7, Doc.\,4, p.\,38)} is, if not theoretically,
but practically excluded.} 
\end{quote}

But the \emph{principle of general covariance} is intended as a non-trivial 
selection criterion. Hence modern writers often characterise it as the 
requirement of diffeomorphism invariance, i.e. that the diffeomorphism 
group of spacetime is a symmetry group in the sense of 
Definition\,\ref{def:SymmetryGroup}. But then, as we have seen above, 
the principle is open to trivialisations if one allows background 
structures to become formally dynamical. This possibility can only 
be inhibited if one limits the amount of structure that may be added 
to the dynamical fields.\footnote{Physically speaking, one may be 
tempted to just disallow such formal `equations of motions' whose 
solution space is (up to gauge equivalence) zero dimensional. 
But this would mean that one would have to first understand the 
solution space of a given theory before one can decide on its 
`general covariance' properties, which would presumably render 
it a practically fairly useless criterion.} 

The reason why I mention all this here is that Pauli's 
Relativity article is, to my knowledge, the only one that seems to 
address that point, albeit not as explicitly as one might wish. 
After mentioning Kretschmann's objection, he remarks (the emphases
are Pauli's):%
\footnote{German original:
Einen physikalischen Inhalt bekommt die allgemein kovariante Formulierung 
der Naturgesetze erst durch das \"Aquivalenzprinzip, welches zur Folge hat, 
da\ss\ die Gravitation durch die $g_{ik}$ \emph{allein} beschrieben wird, und 
da\ss\ diese nicht unabh\"angig von der Materie gegeben, sondern selbst 
durch Feldgleichungen bestimmt sind. Erst deshalb k\"onnen die $g_{ik}$ 
als \emph{physikalische Zustandsgr\"o{\ss}en} bezeichnet werden. 
(\cite{Pauli:2000}, p.\,181)}
\begin{quote}
\emph{
The generally covariant formulation of the physical laws acquires a physical content 
only through the principle of equivalence, in consequence of which gravitation is 
described \emph{solely} by the $g_{ik}$ and the latter are not given 
independently from matter, but are themselves determined by the field 
equations. Only for this reason can the $g_{ik}$ be described as 
\emph{physical quantities}.} 
(\cite{Pauli:ToR-Dover}, p.\,150)
\end{quote}
Note how perceptive Pauli addresses the two central issues:
1)~that one has to limit the the amount of dynamical variables and
2)~that dynamical structures have to legitimate themselves as 
physical quantities through their back reaction onto other 
(matter) structures. It is by far the best few-line account of 
the issue that I know of, though perhaps a little hard to 
understand without the more detailed discussion given above in 
Section\,\ref{sec:DynamicalSymmetriesVersusCovariance}. 
Most modern textbooks do not even address the problem. 
See \cite{Giulini:2007b} for more discussion.

\subsection{General covariance and antimatter}
\label{sec:Antimatter}
In this section I wish to give a brief but illustrative example
from Pauli's work for the non-trivial distinction between 
observable physical symmetries on one hand, and gauge symmetries 
on the other (cf. Section\,\ref{sec:PhysicalVersusGaugeSymmetries}). 
The example I 
have chosen concerns an argument within the (now outdated) attempts 
to understand elementary particles as regular solutions of classical 
field equations. Pauli reviewed such attempts in a rather detailed 
fashion in his Relativity article, with particular emphasis on 
Weyl's theory, to which he had actively contributed in two of 
his first three published papers in 1919. 

The argument proper 
says that in any `generally covariant'\footnote{Here `general 
covariance' is taken to mean that the diffeomorphism group of 
spacetime acts as symmetry group.} theory, which allows for 
regular static solutions representing charged particles, there 
exists for any solution with mass $m$ and charge $e$ another 
such solution with the same mass but opposite charge $-e$. 
Pauli's proof looks like an almost trivial application of 
general covariance and runs as follows: Let $g_{\mu\nu}(x^\lambda)$ 
and $A_\mu(x^\lambda)$ represent the gravitational and 
electromagnetic field respectively. The hypothesis of 
staticity implies that coordinates (and gauges for $A_\mu$) 
can be chosen such that all fields are independent of the 
time coordinate, $x^0$, and that $g_{0i}\equiv 0$ as well as 
$A_i\equiv 0$ for $i=1,2,3$.\footnote{The latter conditions 
distinguish staticity from mere stationarity. The condition 
on $A_i$ may, in fact, be relaxed.} Now consider the 
orientation-reversing diffeomorphism 
$\phi:(x^0,\vec x)\mapsto (-x^0,\vec x)$. 
It maps the gravitational field to itself while reversing the 
sign of $A_0$ and hence of the electric field. General covariance 
assures these new fields to be again solutions with the same total 
mass but opposite total electric charge. 

Pauli presents this argument in his second paper addressing
Weyl's theory, entitled \emph{To the Theory of Gravitation
and Electricity by Hermann Weyl}\footnote{German original: 
``Zur Theorie der Gravitation und der Elektrizit\"at von Hermann 
Weyl''.} (\cite{Pauli:CSP}, Vol.\,2, pp.\,13-23, here p.\,18) 
and also towards the end of Section\,67 of his Relativity 
article. The idea of this proof is due to Weyl who communicated it 
(without formulae) in his first two letters to Pauli 
(\cite{Pauli:SC}, Vol.\,1, Doc.\,[1] and [2]), as Pauli also 
acknowledges in his paper (\cite{Pauli:CSP}, Vol.\,2, p.\,18, 
footnote\,2). 

It is interesting to note that Einstein rediscovered the very 
same argument in 1925 and found it worthy of a separate 
communication~\cite{Einstein:1925}. At the time it was common 
to all, Weyl, Pauli, and Einstein, to regard the argument a 
nuisance and of essentially destructive nature. This was because 
at this time antiparticles had not yet been discovered so that 
the apparent asymmetry as regards the sign of the electric 
charges of fundamental particles was believed to be a fundamental
property of Nature. Already in his first paper on Weyl's theory
(\cite{Pauli:CSP}, Vol.\,2, pp.\,1-9), entitled 
\emph{Perihelion Motion of Mercury and Deflection of Rays in Weyl's Theory of 
Gravitation}\footnote{German original: ``Merkurperihelbewegung und 
Strahlenablenkung in Weyls Gravitationstheorie''.}, Pauli 
emphasised:\footnote{German original: 
``Die Hauptschwierigkeit ist -- neben Einstein's Einwand, der 
mir durchaus noch nicht hinreichend widerlegt scheint --, da\ss\ 
die Theorie von der Asymmetrie der beiden Elektrizit\"atsarten 
nicht befriedigend Rechenschaft zu geben vermag.'' 
(\cite{Pauli:CSP}, Vol.\,2, p.\,8)}
\begin{quote}
\emph{
The main difficulty [with Weyl's theory] is -- apart from 
Einstein's objection, which appears to me not yet 
sufficiently disproved -- that the theory cannot account 
for the asymmetry between the two sorts of electricity.}
\end{quote}

Now, there is an interesting conceptual point hidden in this argument 
that relates to our discussions in 
Sections\,\ref{sec:DynamicalSymmetriesVersusCovariance} and 
\ref{sec:PhysicalVersusGaugeSymmetries}. First of all, the two 
solutions are clearly considered physically distinct, otherwise 
the argument could not be understood as contradicting the 
charge asymmetry in Nature. Hence the diffeomorphism involved 
cannot be considered a gauge transformation but rather
corresponds to a proper physical symmetry. On the other hand, 
we know that diffeomorphisms within bounded regions must be 
considered as gauge transformations, for otherwise one would run 
into the dilemma set by the so-called 
``hole argument''\footnote{Let $\Omega$ be a bounded region in 
spacetime which is disjoint from a spacelike hypersurface 
$\Sigma$. Consider two solutions to the field equations which 
merely differ by the action of a diffeomorphism $\phi$ with 
support in $\Omega$. If they are considered distinct, 
then the theory cannot have a well posed initial-value problem, 
since then for any $\Sigma$ distinct solutions exist with 
identical data on $\Sigma$. This is a rephrasing of Einstein's 
original argument (\cite{Einstein:CP}, Vol.\,4, Doc.\,25, p.\,574, 
Doc.\,26, p\,580, Vol.\,6, Doc.\,2, p.\,10), which did not construct 
a contradiction to the existence of a well posed \emph{initial-value 
problem}, but rather to the requirement that the gravitational field 
be determined by the matter content (more precisely: its energy 
momentum tensor). But this requirement is clearly never fulfilled 
in any generally covariant theory in which the gravitational 
field has its own degrees of freedom, independent of whether one 
regards diffeomorphisms as gauge. Slightly later he rephrased it 
so as to construct a contradiction to the existence of a well posed 
\emph{boundary-value problem} (\cite{Einstein:CP}, Vol.\,6, Doc.\,9, p.\,110), 
which is also not the right thing to require from equations that 
describe the propagation of fields with own degrees of freedom.}. Hence one 
faces the problem of how one should characterise those diffeomorphisms
which are not to be considered as gauge transformations 
(cf. Section\,\ref{sec:PhysicalVersusGaugeSymmetries}). It is conceivable 
that this question is not decidable without contextual information.
(See e.g. \cite{Giulini:1995d} and Chapter\,6 of \cite{Joos-etal:2003} 
for more discussion of this point.) 
The historical sources have almost nothing to say about this, though  
there are suggestions by all three mentioned authors how to circumvent 
the argument by adding more non-dynamical structures, as a result of 
which general covariance is lost. Einstein, being most explicit here, 
suggested the existence of a timelike vector field which 
fixes a time orientation. At least the so-defined time orientation 
would then have to be considered as non-dynamical structure of type $\Sigma$ 
(cf. Section\,\ref{sec:DynamicalSymmetriesVersusCovariance}) in 
order to break the symmetry group down to the stabiliser group of
$\Sigma$. The time-orientation-reversing transformation used above 
would then not be a symmetry anymore. Similar suggestions were made 
by Weyl, who also hinted at a structure to distinguish past and
future:\footnote{German original: ``Ihren Wesensunterschied 
[von Vergangenheit und Zukunft] halte ich, im Gegensatz zu den 
meisten Physikern, f\"ur eine Tatsache von noch viel 
fundamentalerer Bedeutung als der Wesensunterschied zwischen 
positiver und negativer Elektrizit\"at.'' (\cite{Pauli:SC}, 
Vol.\,1, Doc.\,[2], p.\,6)}   
\begin{quote}
\emph{
Their essential difference [of past and future] I take, contrary to 
most physicists, to be a fact of much more fundamental meaning than 
the essential difference between positive and negative charge.}
\end{quote} 
In the last (5th) edition of \emph{Raum Zeit Materie}, Hermann Weyl
writes regarding his unified theory (the emphases are Weyl's):
\footnote{German original: ``Die Theorie gibt keinen Aufschlu{\ss}
\"uber die \emph{Ungleichartigkeit von positiver und negativer 
Elektrizit\"at}. Das kann ihr aber nicht zum Vorwurf gemacht werden. 
Denn jene Ungleichartigkeit beruht ohne Zweifel darauf, da{\ss} von 
den beiden Urbestandteilender der Materie, Elektron und Wasserstoffkern, 
der positiv geladene mit einer anderen Masse verbunden ist als der 
negaiv geladene; sie entspringt aus der Natur der Materie und nicht des
Feldes.'' (\cite{Weyl:RZM1991}, p.\,308)}
\begin{quote}
\emph{
The theory gives no clue as regards the \emph{disparity of positive and negative
electricity}. But that cannot be taken as a reproach against the theory. For
that disparity is based without doubt on the fact that of both fundamental 
constituents of matter, the electron and the hydrogen nucleus, the positively 
charged one is tight to another mass then the negatively charged one;
it originates from the nature of matter and not of the field.}
\end{quote}
Given that Weyl is talking about his unified field-theory of gravity
and electricity, whose original claim was to explain all of matter 
by means of field theory, this statement seems rather surprising. It may 
be taken as a sign of Weyl's beginning retreat from his once so 
ambitious programme.

\subsection{Missed opportunities}
\label{sec:MissedOpportunities}
\subsubsection{Supersymmetry}
One issue that attracted much attention during the 1960s was, whether 
the observed particle multiplets could be understood on the basis of
an all embracing symmetry principle that would combine the
Poincar\'e group with the internal symmetry groups displayed by
the multiplet structures. This combination of groups should be non-trivial,
i.e., not be a direct product, for otherwise the internal symmetries would
commute with the spacetime symmetries and lead to multiplets degenerate 
in mass and spin (see, e.g., \cite{Raifeartaigh:1965c}). Subsequently, a 
number of no-go theorems appeared, which culminated in
the now most famous theorem of Coleman \& 
Mandula~\cite{Coleman.Mandula:1967}. This theorem
states that those generators of symmetries of the $S$-matrix
belonging to the Poincar\'e group necessarily commute with those
belonging to internal symmetries. The theorem is based on a series of
assumptions\footnote{The assumptions are:
(1)~there exists a non-trivial (i.e., $\not =\mathbf{1}$) S-matrix
   which depends analytically on $s$ (the squared centre-of-mass
   energy) and $t$ (the squared momentum transfer);
(2)~the mass spectrum of one-particle states consists of (possibly
   infinite) isolated points with only finite degeneracies;
(3)~the generators (of the Lie algebra) of symmetries of the $S$-matrix
   contains (as a Lie-sub algebra) the Poincar\'e generators;
(4)~some technical assumptions concerning the possibility of
   writing the symmetry generators as integral operators in
   momentum space.}
involving the crucial technical condition that the $S$-matrix depends 
analytically on standard scattering parameters. What is less visible 
here is that the structure of the Poincar\'e group enters in a decisive 
way. This result would not follow for the Galilean group, as was explicitly 
pointed out by Coleman \& Mandula (\cite{Coleman.Mandula:1967}, p.\,159).

One way to avoid the theorem of Coleman \& Mandula is to generalise
the notion of symmetries. An early attempt was made by Golfand \&
Likhtman~\cite{Golfand.Likhtman:1971}, who constructed what is now 
known as a Super-Lie algebra, which generalises the concept of Lie 
algebra (i.e. symmetry generators obeying certain commutation relations) 
to one also involving anti-commutators. In this way it became possible
for the first time to link particles of integer and half-integer spin
by a symmetry principle. It is true that Supersymmetry still maintains
the degeneracy in masses and hence cannot account for the mass
differences in multiplets. But its most convincing property,
the symmetry between bosons and fermions, suggested a most elegant
resolution of the notorious ultraviolet divergences that beset
Quantum Field Theory.

It is remarkable that the idea of a cancellation of bosonic and
fermionic contributions to the vacuum energy density occurred to
Pauli. In his lectures \emph{Selected Topics in Field Quantization},
delivered in 1950-51 (in print again since 2000, 
\cite{Pauli:LecturesFieldQuant}), he posed the question
\begin{quote}
\emph{%
 ..whether these zero-point energies [from Bosons and Fermions] can compensate each other.}
(\cite{Pauli:LecturesFieldQuant}, p.\,33)
\end{quote}
He tried to answer this question by writing down the formal
expression for the zero point energy density of a quantum field 
of spin $j$ and mass $m_j>0$ (Pauli restricted attention to spin 
$0$ and spin $1/2$, but the generalisation is immediate):
\begin{equation}
\label{eq:PauliVenergy1}
4\pi^2\frac{E_j}{V}=
(-1)^{2j}(2j+1)\int dk\,k^2\sqrt{k^2+m^2}\,.
\end{equation}
Cancellation should take place for high values of $k$.
The expansion
\begin{equation}
\label{eq:PauliVenergy2}
4\int_0^Kdk\,k^2\sqrt{k^2+m^2}=
K^4+m_j^2K^2-m_j^4\log(2K/m_j)
+O(K^{-1})
\end{equation}
shows that the quartic, quadratic, and logarithmic terms must
cancel in the sum over $j$ for the limit $K\rightarrow\infty$
to exist. This implies that for $n=0,2,4$ one must have
\begin{equation}
\label{eq:PauliVenergy3}
\sum_j(-1)^{2j}(2j+1)m_j^n=0\quad \text{and}\quad
\sum_j(-1)^{2j}(2j+1)\log(m_j)=0\,.
\end{equation}
Pauli comments that 
\begin{quote}
\emph{%
these requirements are so extensive that it is rather improbable that they 
are satisfied in reality.} (\cite{Pauli:LecturesFieldQuant}, p.\,33)
\end{quote}
\emph{Unless enforced by an underlying symmetry}, one is tempted to 
add! This would have been the first call for a supersymmetry in 
the year 1951.  

However, the real world does not seem to be as simple as that. 
Supersymmetry, if at all existent, is strongly broken in the 
phase we live in. So far no supersymmetric partner of any existing 
particle has been detected, even though some of them (e.g., the 
neutralino) are currently suggested to be viable candidates for 
the missing-mass problem in cosmology. Future findings (or 
non-findings) at the Large Hadron Collider (LHC) will probably 
have a decisive impact on the future of the idea of supersymmetry, 
which---whether or not it is realised in Nature---is certainly 
very attractive; and Pauli came close to it.

\subsubsection{Kaluza-Klein Monopoles}
Ever since its first formulation in 1921, Pauli as well as Einstein were 
much attracted by the geometric idea of Theodor Kaluza and its refinement 
by Oskar Klein, according to which the classical theories of the 
gravitational and the electromagnetic field could be unified into a 
single theory, in which the unified field has the same meaning as 
Einsteins gravitational field in General Relativity, namely as metric 
tensor of spacetime, but now in five instead of four dimensions. 
The momentum of a particle in the additional fifth direction 
(which is spacelike) is now to be interpreted as its charge. Charge is 
conserved because the geometry of spacetime is \emph{a priori} restricted to be 
independent of that fifth direction. The combined field equations are 
exactly the five-dimensional analog of Einstein's equations for 
General Relativity. 

A natural question to address in this unified classical theory was 
whether it admits solutions that could represent particle-like 
objects. More precisely, the solution should be stationary, everywhere 
regular, and possess long-ranging gravitational and electromagnetic 
fields (usually associated with aspects of mass and charge). Pauli, 
who was very well familiar with this theory since its first 
appearance\footnote{It came out too late to be considered in the 
first edition of Pauli's Relativity article. 
But he devoted to it a comparatively large space in his 
Supplementary Notes written in early 1956 for the first 
English edition (\cite{Pauli:ToR-Dover}, Suppl. Note\,23, pp.\,227-232;
\cite{Pauli:2000}, pp.\,276-282)}, kept an active interest in it 
even after the formulations of Quantum Mechanics and early Quantum 
Electrodynamics, which made it unquestionable for him that the 
problem of matter could not be adequately addressed in the framework 
of a classical field theory, unlike Einstein, who maintained such a 
hope in various forms until the end of his life in 1955. 

It is therefore remarkable that in 1943 Einstein and Pauli wrote a 
paper in which they proved the non-existence of such solutions. 
The introduction contains the following statement:
\begin{quote}
\emph{
When one tries to find a unified theory of the gravitational and electromagnetic fields,
he cannot help feeling that there is some truth in \emph{Kaluza's} five-dimensional theory.
(\cite{Einstein.Pauli:1943}, p.\,131)}
\end{quote} 
In fact, Einstein and Pauli offered a proof for the more general situation 
with an arbitrary number of additional space dimensions, fulfilling the 
generalised Kaluza-Klein ``cylinder-condition'' that the gravitational field should
not depend on any of these extra directions. Note that this extra 
condition introduces non-dynamical background structures, so that of the 
5-dimensional diffeomorphism group only those diffeomorphisms preserving 
this condition can act as symmetries, a point Pauli often emphasised as 
a deficiency regarding the Kaluza-Klein approach. 

Restricting attention to five dimensions, the explicitly stated hypotheses 
underlying the proof were these (\cite{Einstein.Pauli:1943}, p.\,131; 
annotations in square brackets within quotations are mine): 
\begin{itemize}
\item[H1]
``The field is stationary (i.e the $g_{ik}$ [the five-dimensional metric]
are independent of $x^4$ [the time coordinate]).'' Clearly, $g_{ik}$ is 
also assumed to be independent of the fifth coordinate $x^5$.
\item[H2]
``It [the field $g_{ik}$] is free from singularities.''
\item[H3]
``It is imbedded in a Euclidean space (of the Minkowski type),
and for large values of $r$ ($r$ being the distance from the origin
of the spatial coordinate system) $g_{44}$ has the asymptotic form 
$g_{44}=-1+\mu/r$, where $\mu\ne 0$.'' The last condition is meant 
to assure the non-triviality of the solution, i.e. that there really 
is an attracting object at the spatial origin. This becomes clear if
one recalls that in the lowest weak-field and slow-motion approximation 
$1+g_{44}$ just corresponds to the Newtonian gravitational potential. 
Unfortunately, the other statement: ``It is imbedded in a Euclidean 
space (of the Minkowski type)'' seems ambiguous, since the solution 
is clearly not meant to be just (a portion of) 5-dimensional flat 
Minkowski space. Hence the next closest reading is presumably that 
the underlying five-dimensional spacetime manifold is (diffeomorphic 
to) $\mathbb{R}^5$, with some non-flat metric of Minkowskian 
signature $(-,+,+,+,+)$.\footnote{In fact, 
it turns out that formally the proof does not depend on whether the 
fifth dimension is space- or time-like, as noted by Einstein and 
Pauli (\cite{Einstein.Pauli:1943}, p.\,134).}
\end{itemize} 
The elegant method of proof makes essential use of the fact that 
the suitably restricted group of spacetime diffeomorphisms (to
those preserving the cylinder condition) is a symmetry group 
for the full set of equations in the sense of 
(\ref{eq:DefSymmetry}) of Definition\,\ref{def:SymmetryGroup}.  
More precisely, two types of diffeomorphisms from that class are 
considered separately by Einstein and Pauli: 
\begin{itemize}
\item[D1]
Arbitrary ones in the three coordinates $(x^1,x^2,x^3)$ which 
leave invariant the $(x^4,x^5)$ coordinates.
\item[D2]  
Linear ones in the $(x^4,x^5)$ coordinates, leaving 
invariant the $(x^1,x^2,x^3)$.
\end{itemize}

Now, as a matter of fact, this innocent looking split introduces 
a further and, as it turns out, crucial restriction, over and above 
the hypotheses H1-H3. The point is that the split and, in 
particular, the set D2 of diffeomorphisms simply do not exist 
unless the spacetime manifold, which in H3 was assumed to be 
$\mathbb{R}^5$, globally splits into $\mathbb{R}^2\times\mathbb{R}^3$
such that the first factor, $\mathbb{R}^2$, corresponds to 
the $x^4x^5$-planes of constant spatial coordinates 
$(x^1,x^2,x^3)$ and the second factor, $\mathbb{R}^3$,
corresponds to the $x^1x^2x^3$-spaces of constant coordinates 
$(x^4,x^5)$. But this need not be the case if H1-H3 are assumed. 
The identity derived by Einstein and Pauli from the requirement 
that transformations of the field induced by diffeomorphisms 
of the type D2 are symmetries are absolutely crucial in proving 
the non-existence of regular solutions.\footnote{Specifically 
we mean their identity (13), which together with spatial 
regularity implies the integral form (13a), which in turn leads 
directly to vanishing mass in (22-23a).  (All references are to 
their formulae in \cite{Einstein.Pauli:1943}.)} 

We now know that this additional restriction is essential to the 
non-existence result: There do exist solutions of the type envisaged 
that satisfy H1-H3, but violate the extra (and superfluous) splitting 
condition.\footnote{The somewhat intricate topology of the Kaluza-Klein 
spacetime is this: The $x^5$ coordinate parametrises circles which 
combine with the 2-spheres (polar coordinates $(\theta,\varphi)$) of 
constant spatial radius, $r$, into 3-spheres (Hopf fibration) which 
are parametrised by $(\theta,\varphi,x^5)$, now thought of as Euler 
angles. The radii of these 3-spheres appropriately shrink to zero as 
$r$ tends to zero, so that $(r,\theta,\varphi,x^5)$ define, in fact, 
polar coordinates of $\mathbb{R}^4$. Together with time, $x^4$, we 
get $\mathbb{R}^5$ as global topology. Now the submanifolds of 
constant $(x^1,x^2,x^3)$ are  those of constant $(r,\theta,\varphi)$ 
and have a topology $\mathbb{R}\times S^1$ rather than $\mathbb{R}^2$, 
so that the linear transformations D2 in the $x^4x^5$ coordinates 
do not define diffeomorphisms of the Kaluza-Klein spacetime manifold.} 
They are called \emph{Kaluza-Klein Monopoles} 
\cite{Sorkin:1983}\cite{Gross.Perry:1983} and carry a 
gravitational mass as well as a magnetic charge. It is hard to 
believe that Pauli as well as Einstein would not have been 
much impressed by those solutions, though possibly with different 
conclusions, had they ever learned about them. It is also conceivable 
that these solutions could have been found at the time, had real 
attempts been made, rather than---possibly---discouraged by Pauli's 
and Einstein's result. In fact, Kurt G\"odel, who was already in 
Princeton when Pauli visited Einstein, found his famous 
cosmological solution~\cite{Goedel:1949} in 1949 by a very 
similar geometric insight that also first led to the 
Kaluza-Klein monopole~\cite{Sorkin:1983}.\footnote{Both use 
invariant metrics on 3-dimensional group manifolds, 
$SU(2)$ in the KK case, $SU(1,1)$ in G\"odels case. This 
simplifies the calculations considerably.}

\subsection{Irritations and psychological prejudices}
\label{sec:Irritations}
One of Pauli's major interests were discrete symmetries, in 
particular the transformation of space inversion, 
$\vec x\mapsto -\vec x$, also called \emph{parity} transformation. 
Given a linear wave equation which is symmetric under the proper 
orthochronous (i.e. including no space and time inversions) 
Poincar\'e group, one may ask whether it is also symmetric under 
space and time inversions. For this to be a well defined question 
one has to formulate conditions on how these inversions interact 
with Poincar\'e transformations. Let us focus on the operation 
of space inversion. If this operation is implementable by an 
operator $\Parity$, it must conjugate each rotation and each 
time translation to their respective self, and each boost 
and each space translation to their respective inverse. 
This follows simply from the geometric meaning of space inversion. 
Hence, generally speaking, we need to distinguish the following 
three possible scenarios (recall the notation from 
Section\,\ref{sec:DynamicalSymmetriesVersusCovariance}): 
\begin{itemize}
\item[(a)]
 $\Parity$ acts on $\KPT$ and is a symmetry, i.e. leaves 
 $\DPT_\Sigma\subset\KPT$ invariant; 
\item[(b)]
 $\Parity$ acts on $\KPT$ and is no symmetry, i.e. leaves 
 $\DPT_\Sigma\subset\KPT$ not invariant;
\item[(c)]
$\Parity$ is not implementable on $\KPT$. 
\end{itemize}
It is clear that when one states that a certain equation is not 
symmetric under $\Parity$ one usually addresses situation (b), 
though situation (c) also occurs, as we shall see.  

Consider now the field of a massless spin-$\tfrac{1}{2}$ 
particle, that transforms irreducibly under the proper 
orthochronous Poincar\'e group. The field is then either a 
two-component spinor, $\phi^A$, which in the absence of 
interactions obeys the so-called Weyl equation\footnote{%
Here I use the standard Spinor notation where upper-case 
capital Latin indices refer to (components of) elements in 
spinor space (2-dimensional complex vector space), lower case 
indices to the dual space, and primed indices to the respective 
complex-conjugate spaces. Indices are raised and lowered by 
using a (unique up to scale) $\mathrm{SL}(2{,}\mathbb{C})$
invariant 2-form. An overbar denotes the map into the 
complex-conjugate vector space. Unless stated otherwise, my 
conventions are those of~\cite{Sexl.Urbantke:RGP}.} 
\begin{equation}
\label{eq:WeylEq}
\partial_{AA'}\phi^A=0\,.
\end{equation}
Alternatively, one may also start from a four-component Dirac
spinor, 
\begin{equation}
\label{eq:DiracSpinor}
\psi=
\begin{pmatrix}
\phi^A\\
\bar\chi_{A'}
\end{pmatrix}
\end{equation}
which carries a reducible representation of the proper orthochronous 
Poincar\'e group: If $\phi^A$ transforms with $A\in SL(2{,}\mathbb{C})$ 
then $\bar\chi_{A'}$ transforms with $(A^\dagger)^{-1}$ (being an 
element of the complex-conjugate dual space), so that the space of
the upper two components $\phi^A$ of $\psi$ and the space of the lower 
two components $\bar\chi_{A'}$ of $\psi$ are separately invariant. 
One may then eliminate two of the four components by the so-called 
Majorana condition, which requires the state $\psi$ to be identical 
to its charge-conjugate, $\psi^c$, where
\begin{equation}
\label{eq:CC-DiracSpinor}
\CC: \psi\mapsto\psi^c:=i\gamma^2\psi^*=
\begin{pmatrix}
\chi^A\\
\bar\phi_{A'}
\end{pmatrix}\,.
\end{equation}
Hence for a Majorana spinor one has $\phi=\chi$ and the 
interaction-free Dirac equation reads 
\begin{equation}
\label{eq:DiracEq}
\gamma^\mu\partial_\mu\psi:=
\sqrt{2}\begin{pmatrix}
0&\partial^{AA'}\\
\partial_{A'A}&0
\end{pmatrix}
\begin{pmatrix}
\phi^A\\
\bar\phi_{A'}
\end{pmatrix}
=0\,.
\end{equation}
One can now either regard (\ref{eq:WeylEq}) or (\ref{eq:DiracEq}) 
as the interaction-free equation for a neutrino.  

Here I wish to briefly recall a curious discussion between Pauli
and Fierz on whether or not these two equations describe physically 
different state of affairs. Superficially this discussion is 
about a formal and, mathematically speaking, rather trivial point.
But, as we will see, it relates to deep-lying preconceptions in 
Pauli's thinking about issues of symmetry. This makes it worth 
looking at this episode in some detail. 

First note that there is an obvious bijection, $\beta$, between 
two-component spinors and Majorana spinors, given by
\begin{equation}
\label{eq:SpinorBijection}
\beta:\phi^A\mapsto  
\begin{pmatrix}
\phi^A\\
\bar\phi_{A'}
\end{pmatrix}\,.
\end{equation}
Note also that the set of Majorana spinors is \emph{a priori} a 
\emph{real}\footnote{The reality structure on the complex
vector space of Dirac spinors is provided by the charge 
conjugation map.} vector space, though it has a complex 
structure, $j$, given by 
\begin{equation}
\label{eq:MajoranaCompexStructure}
j:
\begin{pmatrix}
\phi^A\\
\bar\phi_{A'}
\end{pmatrix}
\mapsto
\begin{pmatrix}
i\phi^A\\
-i\bar\phi_{A'}
\end{pmatrix}\,,
\end{equation}
with respect to which the bijection (\ref{eq:SpinorBijection}) 
satisfies $\beta\circ i=j\circ\beta$,  where here $i$ stands for 
the standard complex structure (multiplication with imaginary 
unit $i$) in the space $\mathbb{C}$ of two-component spinors. 
However, regarded as a map between complex vector spaces, the 
bijection $\beta$ is \emph{not} linear.   

Now, Pauli observed already in 1933 (see quotation below) that the 
Weyl equation (\ref{eq:WeylEq}) is not symmetric under parity. 
Hence he concluded it could not be used to describe Nature.
In fact, what is actually the case is that parity cannot even 
be implemented as a \emph{linear} map on the space of two-component 
spinors (case (c) above). This is easy to see and in fact true for 
any irreducible representation of the Lorentz group that stays 
irreducible if restricted to the rotation group (i.e. for purely 
primed or purely unprimed spinors).\footnote{%
\label{foot:ParityImplementability}
As stated above, the geometric meaning of space inversion requires 
that the parity operator (if existent) commutes with spatial 
rotations and conjugates boosts to their inverse. The first requirement 
implies (via Schur's Lemma) that it must be a multiple of the 
identity in any irreducible representation that stays irreducible 
when restricted to the rotation subgroup, which contradicts the 
second requirement. Hence it cannot exist in such representations, 
which are precisely those with only unprimed or only primed indices.} 

On the other hand, the Dirac equation \emph{is} symmetric under 
space inversions. Indeed, the spinor-map corresponding to the 
inversion in the spatial plane perpendicular to the timelike 
normal $n$ is given by 
\begin{equation}
\label{eq:DiracParity}     
\Parity:\psi\mapsto\psi^p:=\eta\,n_\mu\gamma^\mu(\psi\circ\rho_n)\,,
\end{equation}
where $\rho_n:x^\mu\mapsto -x^\mu+2n^\mu(n_\nu x^\nu)$ and where 
$\eta$ is a complex number of unit modulus, called the 
\emph{intrinsic parity} of the particular field $\psi$. It is easy to 
see that $\Parity$ is a symmetry of (\ref{eq:DiracEq}) for any 
$\eta$. Note that $\Parity^2=\eta^2\mathbf{1}$ so that 
$\eta\in\{1,-1,i,-i\}$, since for spinors one only requires 
$\Parity^2=\pm\mathbf{1}$ (rather than $\Parity^2=\mathbf{1}$). 
It is also easy to verify that $\Parity$ commutes with $\CC$ iff
$\eta=\pm i$. So if we assign imaginary parity to the Majorana 
field\footnote{Which is also the standard choice in QFT; see e.g. 
\cite{Weinberg:QToF1}, pp.\,126,226.}, the operator $\Parity$ 
also acts on the subspace of Majorana spinors. We conclude that 
the free Majorana equation \emph{is} parity invariant. 

Hence it seems at first that the Weyl formulation and the 
Majorana formulation differ since they have different symmetry 
properties. But this is not true. Using the bijection 
(\ref{eq:SpinorBijection}), we can pull-back the parity map 
(\ref{eq:DiracParity}) to the space of two-component spinors, 
where it becomes (now either $\eta=i$ or $\eta=-i$)
\begin{equation}
\label{eq:WeylParity}     
\phi^A\mapsto\ \eta\sqrt{2}\ n^{AA'}(\bar\phi_{A'}\circ\rho_n)\,,
\end{equation}
which is now an anti-linear map on the space of two-component 
spinors.   

All this was essentially pointed out to Pauli by Markus Fierz 
in a letter dated February 6th 1957 (\cite{Pauli:SC} Vol.\,IV, 
Part\,IV\,A, Doc.\,[2494], p.\,171) in connection with Lee's and 
Yang's two-component theory of the neutrino. Fierz correctly 
concluded from this essential equivalence\footnote{Meaning the 
existence of a bijection that maps all quantities of interest 
(states, currents, symmetries) of one theory to the other.} that 
the 2-component theory as such (i.e. without interactions) did 
not warrant the conclusion of parity violation; only 
interactions could be held responsible for that. 

This was a relevant point in the theoretical discussion at the time, 
as can be seen from the fact that there were two independent papers 
published in \emph{The Physical Review} shortly after Fierz's private 
letter to Pauli, containing the very same observation. The first 
paper was submitted on February 13th by McLennan~\cite{McLennan:1957}, 
the second on March 25th by Case~\cite{Case:1957}. In fact, 
Serpe made this observation already in 1952 \cite{Serpe:1952} 
and emphasised it once more in 1957~\cite{Serpe:1957}. 

One might be worried about the anti-linearity of the transformation
in (\ref{eq:WeylParity}). In that respect, also following Fierz, an 
illuminating analogy may be mentioned regarding the vacuum Maxwell 
equations, which can be written in the form 
\begin{equation} 
\label{eq:ComplexMaxwellEq1}
i\partial_t\vec\Phi-\vec\nabla\times\vec\Phi=0\,,\qquad
\vec\nabla\cdot\vec\Phi=0\,,
\end{equation}
where 
\begin{equation} 
\label{eq:ComplexMaxwellEq2}
\vec\Phi:=\vec E+i\vec B
\end{equation}
is a complex combination of the electric and magnetic field. 
Both equations (\ref{eq:ComplexMaxwellEq1}) are
clearly equivalent of the full set of Maxwell's equations. 
It can be shown that spatial inversions cannot be implemented 
as complex-linear transformations on the complex-valued field 
$\vec\Phi$.\footnote{Equations (\ref{eq:ComplexMaxwellEq1})
are equivalent to $\partial^{AA'}f_{AB}=0$, where $f_{AB}$
is the unprimed spinor equivalent of the tensor $F_{\mu\nu}$ for 
the electromagnetic field strength. Parity cannot be linearly 
implemented on this purely unprimed spinor, for reasons already 
explained in footnote\,\ref{foot:ParityImplementability}.} 
But, clearly, we 
know that Maxwell's equations
are parity invariant, namely if we transform the electric field
as $\vec E\mapsto -\vec E\circ\rho$ (`polar' vector-field)
and the magnetic field as $\vec B\mapsto \vec B\circ\rho$ (`axial'
vector-field), where $\rho: (t,\vec x)\mapsto (t,-\vec x)$. 
This corresponds to an \emph{anti}linear symmetry of 
(\ref{eq:ComplexMaxwellEq1}), given by 
$\vec\Phi\mapsto -\bar{\vec\Phi}\circ\rho$.

Coming back to Fierz's (and other's) original observation for the 
spinor field, they were accepted without much ado by others. 
For example, in her survey on the neutrino in the Pauli 
Memorial Volume, Madame Wu states that ``It is the interaction and 
the interaction only that violates parity'' 
(\cite{Fierz.Weisskopf:PauliMemorial}, footnote on p.\,270.).
In note 25c of that paper she explicitly thanks Fierz for 
``enlightening discussions'' on the two-component theory of the 
neutrino. Clearly Fierz expected his observation to be of interest 
to Pauli, who had already in the 1933 first 
edition of his handbook article on wave mechanics propagated the 
view that Weyl's two-component equations are\footnote{German 
original, full sentence: ``Indessen sind diese
Wellengleichungen, wie ja aus ihrer Herleitung hervorgeht, nicht 
invariant gegen\"uber Spiegelungen (Vertauschung von links und rechts) 
und infolge dessen sind sie auf die physikalische Wirklichkeit nicht 
anwendbar'' (\cite{Pauli:Wellenmechanik}, p.\,234, note 54).
The conclusion concerning non applicability to the physical reality
is cancelled in the 1958 edition; cf. \cite{Pauli:Wellenmechanik}, p.\,150.}
\begin{quote}
\emph{
...not invariant under reflections (interchange of left and right) and,
as a consequence, not applicable to the physical reality.}
\end{quote} 
But instead, Pauli reacts with a surprising plethora of 
ridiculing remarks:\footnote{%
German original: ``Lieber Herr Fierz! Ihr Brief vom 6.
ist der gr\"o{\ss}te Bock den Sie im Laufe Ihres Lebens geschossen 
haben! (Wahrscheinlich kommt heute Nachmittag schon eine Berichtigung
von Ihnen.) Habe nur den ersten Absatz Ihres der Anstalt entsprungenen 
Briefes gelesen und mich gesch\"uttelt vor lachen. [...]
Wenn dieser Brief ankommt (Ihren rahme ich ein!), wissen Sie
wohl schon alles!'' (\cite{Pauli:SC}, Vol.\,IV, Part\,IV\,A, 
Doc.\,[2497], p.\,179).}
\begin{quote}
\emph{%
Dear Mr. Fierz! Your letter from the 6th is the biggest blunder you ever commited
in your life! (Probably this afternoon you will send a correction). Have only read 
the first paragraph of your letter which originated in the asylum and was shaking 
with laughter. [...] When this letter arrives (yours I will frame!) you probably 
will already know everything.}
\end{quote}
Personal irritations emerged which lasted about one week through 
several exchanges of letters and a phone-call. Finally Pauli essentially 
conceded Fierz's point in a long letter of February 12th 1957 that 
also contains first hints at Pauli's psychological resistances
(the emphasis is Pauli's):%
\footnote{German Original: ``Ihre Darstellung erzeugt bei mir das 
Gef\"uhl der `formalistischen Langeweile'', zu der die Lachsalve
\emph{kompensatorisch} war'' (\cite{Pauli:SC},  Vol.\,IV, Part\,IV\,A, 
Doc.\,[2510], p.\,197).} 
\begin{quote}
\emph{
Your presentation creates in me a feeling of ``formal boredom'', to
which the fusillade of laughter was of a \emph{compensatory} nature.}
\end{quote}
This is a curious episode and not easy to understand.  
Pauli's point seems to have been that he wanted to maintain the 
particle-antiparticle distinction \emph{independently} of parity, 
whereas Fierz pointed out that the two-component theory provided 
no corresponding structural element: In Weyl's form the operations 
$\CC$ and $\Parity$ simply do not exist separately, in the Majorana form 
$\Parity$ exists and $\CC$ is the identity (hence not 
distinguishing). Psychologically speaking, Pauli's point becomes 
perhaps more understandable if one takes into account the fact 
that since the fall of 1956 he was thinking about the question 
of lepton-charge conservation. Intuitively he had therefore 
taken as self-evident that opposite helicities also corresponded
to the particle-antiparticle duality (cf. \cite{Pauli:SC}, Vol.\,IV, 
Part\,IV\,A, Doc.\,[2497]), even though this mental association 
did not correspond to anything in the equations. In a letter dated 
February 15th 1957 he offered the following in-depth psychological 
explanation to Fierz (the emphases are Pauli's):%
\footnote{German original:
``Also die `Lachsalve' erfolgte beim Wort `Majorana Theorie' Ihres
ersten Briefes, ich konnte nach diesem Stichwort nicht mehr 
weiterlesen. Die unmittelbare Assoziation zu Majorana war 
nat\"urlich `aha, Teilchen und Antiteilchen soll es nicht mehr 
geben, die will man wir wegnehmen (wie man jemandem ein Symbol
wegnimmt)!' Davor habe ich \emph{Angst}. Ich weiss auch, da{\ss}
mir schon seit Herbst die Erhaltung der Leptonladung in der Physik
ungeheuer wichtig ist -- rational betrachtet, vielleicht 
\emph{zu wichtig}. Ich habe \emph{Angst}, sie k\"onnte sich als 
unrichtig herausstellen und, psychologisch gesehen, ist 
`Unzufriedenheit' ein Euphemismus f\"ur Angst. 
Die $CP$ - ($\equiv$ Majorana $P+$ Vertauschung von Elektron und 
Positron) Invarianz ist mir auch wichtig, aber weniger wichtig als 
die Erhaltung der Leptonladung. Es ist \emph{sicher wahr}, da{\ss} 
`mein platonischer Spiegelkomplex angestochen' war. Teilchen und
Antiteilchen sind das Symbol f\"ur jene allgemeine Spiegelung
(wie weit sie speziell \emph{platonisch} ist, dessen bin ich nicht 
sicher). [...] Offenbar hat der `Spiegelungskomplex' bei mir etwas 
mit Tod und Unsterblichkeit zu tun. \emph{Daher} die Angst! W\"are die 
Beziehung zwischen dem schlafenden Spiegelbild und dem Wachenden 
gest\"ort, oder w\"aren sie gar identisch (Majorana), so g\"abe es,
psychologisch gesprochen, weder Leben (Geburt) noch Tod.''    
(\cite{Pauli:SC}, Vol.\,IV, Part\,IV\,A, Doc.\,[2517], p.\,225)}   
\begin{quote}
\emph{Well, the fusillade of laughter occurred with the expression ``Majorana Theory''
of your first letter. After this catchword I could not go on reading. The immediate 
association with Majorana clearly has been this: ``aha, particles and antiparticles 
should no longer exist, these one intends to take away from me (as one takes away 
a symbol from somebody)!'' This causes me \emph{anxiety}. I also know that 
since last fall the conservation of lepton charge in physics was tremendously 
important to me---looked upon rationally probably \emph{too important}.
I am \emph{anxious} it could turn out to be incorrect and, 
psychologically speaking, ``discontentedness'' is a euphemism 
for anxiety. The $CP$ ($\equiv$ Majorana $P+$ exchange between 
electron and positron) invariance is also important to me, but 
less so than the conservation of lepton charge. It is 
\emph{certainly true} that it ``hit upon my Platonic mirror complex''.
Particles and antiparticles are the symbol for that more general 
mirroring (I am not sure to what extent it is particularly 
\emph{platonic}).[...]
Mirroring is also a \emph{gnostic symbol} for life and death. 
There light is \emph{extinguished} at birth and \emph{lightened up}
at death. [..] Obviously,  for me the ``mirroring complex'' 
has something to do with death and immortality. \emph{Hence} the 
anxiety! If the relation between the sleeping mirror image 
and the one awake would be disturbed, or if they would even be 
identical (Majorana), then, psychologically speaking, there  
would neither be life (birth) nor death.}
\end{quote}
Fierz later commented on that episode in a personal letter 
to Norbert Straumann, parts of which are quoted 
in~\cite{Straumann:1992}.

\subsection{$\beta$-Decay and related issues}
\label{sec:CPTandCosmology}

\subsubsection{CPT}
In 1955 a collection of essays by distinguished physicists 
appeared to celebrate Niels Bohr's 70th 
birthdays~\cite{Pauli:NielsBohr70}. Pauli's 
contribution (\cite{Pauli:NielsBohr70}, p.\,30-51) is entitled 
``Exclusion Principle, Lorentz Group and Reflection of 
Space-Time and Charge'', whose introduction contains the 
following remarks:
\begin{quote}
\emph{%
After a brief period of spiritual and human confusion,
caused by provisional restriction to ``Anschaulichkeit'', 
a general agreement was reached following the substitution 
of abstract mathematical symbols, as for instance psi, for concrete 
pictures. Especially the concrete picture of rotation has been 
replaced by mathematical characteristics of the representations 
of the group of rotations in three dimensional space. This group 
was soon amplified to the Lorentz group in the work of Dirac. 
[...] The mathematical group was further amplified by including 
the reflections of space and time. [...]
I believe that this paper also illustrates the fact that 
a rigorous mathematical formalism and epistemological analysis are both 
indispensable in physics in a complementary way in the sense of 
Niels Bohr. While I try to use the former to connect all mentioned 
features of the theory with help of a richer ``fullness'' of plus and 
minus signs in an increasing ``clarity'', the latter makes me aware 
that the final ``truth'' on the subject is still ``dwelling in the 
abyss''.}\footnote{Here Pauli sets the following footnote: ``I refer 
here to Bohr's favourite verses of Schiller: `Nur die F\"ulle f\"uhrt
zur Klarheit / Und im Abgrund wohnt die Wahrheit'''.} 
(\cite{Pauli:NielsBohr70}, p.\,30-31)
\end{quote} 
In some sense this paper of Pauli's can be seen as a follow-up to his 
spin-statistics paper already discussed above, the main difference 
being that Pauli now considers \emph{interacting} fields. Pauli now 
\emph{assumes} (1)~the validity of the spin-statistics correlation for 
interacting fields (for which there was no proof at the time), 
(2)~invariance under (the universal cover of) the proper orthochronous 
Lorentz group 
$\mathrm{SL}(2{,}\mathbb{C})$ (as in the spin-statistics paper), and 
(3)~locality of the interactions (i.e. involving only finitely many 
derivatives). Then Pauli shows that this suffices to derive the 
so-called CPT theorem that states that the combination of charge 
conjugation (C) and spacetime reflection (PT) is a 
symmetry.\footnote{Pauli used a now outdated terminology: instead of 
CPT he uses SR (strong reflection), instead of PT he uses WR (weak 
reflection), and instead of C he uses AC (antiparticle conjugation).
Preliminary versions of the CPT theorem appeared in papers by 
Julian Schwinger (1951)and Gerhard L\"uders (1954) to which Pauli 
refers. Two years after Pauli's 1955 paper Res Jost gave a very 
elegant proof in the framework of axiomatic quantum field 
theory~\cite{Jost:1957}.} 

At the time (1955) Pauli wrote his paper it was not known whether
any of the operations of $C$, $P$, or $T$ would separately \emph{not}
be a symmetry. This changed when in January 1957 through the experiments 
of Madame Wu \emph{et al.}, in which explicit violations of $P$ and $C$ were 
seen in processes of beta-decay, following a suggestion that this 
should be checked by Lee and Yang in mid 1956~\cite{Lee.Yang:1956a}. 
Pauli had still offered a bet that this would not happen on 
January 17th 1957 (the emphases are Pauli's):\footnote{
German original: ``Ich glaube aber \emph{nicht, da{\ss} der Herrgott ein 
schwacher Linksh\"ander} ist und w\"are bereit hoch zu wetten, da{\ss}
das Experiment symmetrische Winkelverteilung der Elektronen 
(Spiegelinvarianz) ergeben wird. Denn ich sehe keine logische 
Verbindung von \emph{St\"arke} einer Wechselwirkung und ihrer 
Spiegelinvarianz.'' (\cite{Pauli:SC}, Vol.\,IV, Part\,IV\,A, 
Doc.\,[2455], p.\,82)} 
\begin{quote}
\emph{
I do \emph{not}  believe \emph{that God is a weak left-hander} and would 
be prepared to bet a high amount that the experiment will show a 
symmetric angular distribution of the electrons (mirror symmetry). 
For I cannot see a logical connection between the \emph{strength} of 
an interaction and its mirror symmetry.}   
\end{quote}
In view of this firm belief in symmetry the following is remarkable:
In his CPT paper Pauli takes great care to writes down the most 
general ultralocal (i.e. no derivatives) four-fermion interaction 
(for the neutron, proton, electron and neutrino), which is \emph{not} 
$P$ invariant. In contains 10 essentially different terms
with ten coupling constants $C_1,\cdots C_{10}$, only the first five
of which are parity invariant (scalars), whereas the other five are 
pseudoscalars, i.e change sign under spatial inversions. Apparently 
this he did just for the sake of mathematical generality without any 
physical motivation, as he explicitly stated in a letter to Madame 
Wu dated January 19th 1957 (the emphases are Pauli's):
\begin{quote}
\emph{
When I considered such formal possibilities in my paper in the 
Bohr-Festival Volume (1955), I did not think that 
this could have something to do with Nature. I considered it merely 
as a mathematical play, and, as a matter of fact, I did not believe 
in it when I read the paper of Yang and Lee. [...]
What prevented me \emph{until now} from accepting this formal possibility
is the question why this restriction of mirroring appears only in the 
`weak' interactions, not in the strong ones. \emph{Theoretically}, I do not see
any interpretation of this fact, which is empirically so well 
established.} (\cite{Pauli:SC}, Vol.\,IV, Part\,IV\,A, 
Doc.\,[2460], p.\,89)
\end{quote}         
Lee and Yang took this possibility more serious: In an appendix 
to their paper they also write down all ten terms for the 
full, parity non-invariant interaction (\cite{Lee.Yang:1956a}, p.\,258),
without any citation of Pauli.
 
Pauli first learnt that the experiments by Madame Wu \emph{et al.} had 
led to an asymmetric angular distribution from a letter by John 
Blatt from Princeton, dated January 15th 1957. There Blatt writes:
\begin{quote}
\emph{
I don't know whether anyone has written you as yet about the sudden death
of parity. Miss Wu has done an experiment with beta-decay 
of oriented Co nuclei which shows that parity is \emph{not}
conserved in $\beta$ decay. [...] We are all rather shaken 
by by the death of our well-beloved friend, parity.}
(\cite{Pauli:SC}, Vol.\,IV, Part\,IV\,A, Doc.\,[2451], p.\,74)
\end{quote} 
Pauli, too, was shocked as he stated in his famous letter to 
Weisskopf dated January\,27/28 1957 (\cite{Pauli:SC}, 
Vol.\,IV, Part\,IV\,A, Doc.\,[2476]). In that very same letter 
Pauli already started speculating how symmetry could be restored by 
letting the constants $C_i$ become dynamical field, scalar fields 
for $i=1,\cdots, 5$ and pseudo-scalar ones for $i=6,\cdots, 10$:%
\footnote{German original:
``Denken wir uns z.B. die Terme mit $C_1,\cdots,C_5$ mit einem Skalarfeld
$\phi(x)$, die Terme mit $C_6,\cdots,C_{10}$ mit einem 
Pseudo-Skalarfeld  $\hat\phi(x)$ multipliziert. F\"ur den Herrgott, 
der das Vorzeichen von $\hat\phi(x)$ umdrehen kann, w\"are eine solche 
Theorie nat\"urlich rechts-links-invariant -- nicht aber f\"ur uns 
sterbliche Menschen, die wir gar nichts wissen \"uber jenes 
hypothetische neue Feld, au{\ss}er da{\ss} es praktisch auf der Erde 
raum-zeitlich konstant (statisch-homogen) ist, und die wir noch kein 
Mittel haben, es zu \"andern.''  
(\cite{Pauli:SC}, Vol.\,IV, Part\,IV\,A, Doc.\,[2476], pp.\,122-123)}
\begin{quote}
\emph{
Let us imagine, for example, the terms with $C_1,\cdots,C_5$ being 
multiplied with a scalar field $\phi(x)$, the terms 
$C_6,\cdots,C_{10}$ multiplied with a pseudo-scalar field 
$\hat\phi(x)$. For God Himself, Who can change the sign of 
$\hat\phi(x)$, such a theory would be left-right-invariant---not 
for us mortal men, however, who do not know anything about that new 
hypothetical field, except that it is practically constant in space and 
time on earth (static-homogeneous), and that we do not yet\footnote{The
``yet'' is incorrectly omitted in the official translation
(\cite{Pauli:SC}, Vol.\,IV, Part\,IV\,A, p.\,126).} 
have any means to change it.}
\end{quote}         
The mechanism envisaged here to restore symmetry is just 
that discussed in 
Section\,\ref{sec:DynamicalSymmetriesVersusCovariance}, 
where non-dynamical backgrounds structures, $\Sigma$, are 
(formally) turned into dynamical quantities, $\Phi$.

\subsubsection{The Pauli group}
As already mentioned, since fall of 1956 Pauli's thinking about
beta-decay was dominated by the lepton-charge conservation.  
In a paper submitted on March 14th 1957, entitled 
\emph{On the Conservation of Lepton Charge} (\cite{Pauli:CSP}, 
Vol.\,2, pp.\,1338-1349), Pauli once more showed his mastery of 
symmetry considerations while keeping everything at the largest 
possible degree of generality. 

He starts by considering the most general ultralocal four-fermion 
interactions (not necessarily preserving parity or lepton charge) in 
which the neutrino field is represented by a Dirac 4-spinor, $\psi$.
For what follows it is convenient to think of the four components 
of $\psi$ as comprising the following four particle states (per 
momentum): a left-handed neutrino, $\psi_L$, a right-handed neutrino, 
$\psi_R$, and their antiparticles $\psi^c_L$ and $\psi^c_R$ 
respectively. Note that this means $\psi^c_{L,R}:=(\psi_{L,R})^c$ 
and that accordingly $\psi^c_L$ is right- and $\psi^c_R$ is left-handed. 
Here we follow the convention of~\cite{Kemmer.etal:1959}. 

Next Pauli considers a four-parameter group of canonical transformations
(i.e. they leave the anticommutation relations between the fermion fields 
invariant) of the neutrino field, henceforth called the \emph{Pauli group},  
whose interpretation will be given below. These transformations define a 
symmetry of the interaction-free equations of motions (assuming a massless 
neutrino throughout), but will generally \emph{not} define a symmetry once 
the interaction is taken into account. Rather, the following 
is true (cf.~\cite{Nishijima:2004}): Suppose that the general 
interaction depends on a finite number of coupling constants $c_i$ 
for $i=1,\cdots, n$ and that the equations of motion follow 
from an action principle with Lagrange density $\Lag{\Sigma}{\Phi}$, 
where $\Sigma$ represents the array of coupling constants 
(we notationally ignore other non-dynamical structures here) 
and $\Phi$ the dynamical fields. Then the Pauli group acts as 
covariance in a slightly stronger sense than 
(\ref{eq:DefCovariance2}), namely so that 
\begin{equation}
\label{eq:PauliSymmetry}
\Lag{g\cdot\Sigma}{g\cdot\Phi}=\Lag{\Sigma}{\Phi}\,.
\end{equation}
This means that on the level of the Lagrange density (or the 
Hamiltonian), and hence in particular at the level of the equations 
of motion, the transformation of the dynamical fields can be 
compensated for by a transformation of the coupling constants. 
A large part of Pauli's paper is actually devoted to the 
determination of that compensating action of the Pauli group 
on the array of coupling constants.      

Next suppose the initial state is chosen to be invariant under 
the Pauli group, i.e. $g\cdot\Phi=\Phi$ for all $g$.  Then 
(\ref{eq:PauliSymmetry}) implies that its evolution with 
interaction parametrised by $\Sigma$ (the array of $c_i$'s) is 
identical to the evolution parametrised by $g\cdot\Sigma$ for 
any $g$. Hence the outcome of the evolution can only depend on 
the $c_i$'s through their Pauli-invariant combinations.%
\footnote{For illustrative 
purposes we argue here as if all fields were classical and obeyed 
classical equations of motion, though Pauli clearly considers 
the quantum theory where the fields become operators. 
The principal argument is the same, though what makes a big 
difference between the classical and the quantum case is that in 
the latter we can more easily ascertain the existence of invariant initial 
states. This is because in quantum theory, assuming there are no 
superselection rules at work, the superposition principle always 
allows us to construct invariant initial states by group-averaging 
any given state over the group (which is here compact, so that the 
averaging is unambiguously defined). Such states would, for
example, appropriately represent physical situations where those 
observables that distinguish between the states in the group orbit 
are not measured, may it be for reasons of practice or of principle.} 
In particular, since the neutrinoless 
double beta-decay simply has no initial neutrino, this reasoning 
can be applied to it. If this lepton-charge-conservation violating 
process is deemed impossible, the corresponding Pauli-invariant 
combination of coupling constants to which the scattering probability 
is proportional\footnote{It will be a quadratic combination in leading 
order of perturbation theory. Explicit calculations had been done by 
Pauli's assistant Charles Enz \cite{Enz:1957}.} must 
vanish. This, in turn, gives the sought-after constraint on the 
possible four-fermion interaction. For (massless) neutrinos in 
Majorana representation Pauli finally arrived at the result that 
either only the left- or the right-handed component enters the 
interaction. It should be added that this clever sort of reasoning 
was shortly before used by Pursey in a less general 
setting~\cite{Pursey:1957} in which the interaction was specialised 
\emph{a priori} to conserve lepton charge.\footnote{In terms of 
the Pauli group, Pursey did not consider the $U(1)$ part.}  
More on the history of the search for the right form of the 
four-fermion interaction may be found in~\cite{Straumann:1992}. 
It should also be mentioned that the possibility of neutrinoless 
double beta-decays is currently still under active experimental 
investigation at the National Gran Sasso Laboratory, where the 
2003-2005 CUORICINO experiment set upper bounds for the Majorana 
mass of the electron neutrino well below one $eV$. The upcoming 
next-generation experiment, CUORE, is designed to lower this 
bound to $5\cdot 10^{-2}\,meV$; compare \cite{Gorla:2008}.  

What is the interpretation of the Pauli group? Mathematically it is 
isomorphic to $U(2)$, the group of $2\times 2$ unitary matrices 
acting on a two-dimensional complex vector space. Here there are 
two such spaces (per 4-momentum) in which it acts: the 
`left-handed subspace' that is spanned by the two left-handed 
components $\psi_L$ and $\psi^c_R$, and the `right-handed subspace' 
that is spanned by the two right-handed components 
$\psi_R$ and $\psi^c_L$. The two actions of $U(2)$ in these spaces 
are complex conjugate to each other (see equation
(\ref{eq:PauliGroupAction})). Usually one thinks of the 
Pauli group as $U(1)\times SU(2)$, which is a double 
cover of $U(2)$, so that the four real parameters are written 
as a phase $\exp(i\alpha)$, parametrising $U(1)$, and two complex 
parameters $a,b$ satisfying $\vert a\vert^2+\vert b\vert^2=1$,
which give three real parameters when split into real and imaginary 
part and which parametrise a 3-sphere that underlies $SU(2)$ as 
group manifold. In this parametrisation the action of the Pauli 
group reads (an asterisk stands for complex conjugation):%
\footnote{Usually the Pauli group is written in terms 
of the 4-component neutrino field $\psi$ as 
$\psi\mapsto \exp(i\alpha\gamma_5)(a\psi+b\gamma_5\psi^c)$,
where $\psi^c:=i\gamma_2\psi^*$ is the charge conjugate field. 
But this is easily seen to be equivalent to (\ref{eq:PauliGroupAction}) 
if one sets $\psi_{L,R}=\frac{1}{2}(1\pm\gamma_5)\psi$ and 
$\psi^c_{R,L}=\frac{1}{2}(1\pm\gamma_5)\psi^c$. The more explicit 
form (\ref{eq:PauliGroupAction}) is better suited for the 
interpretational discussion; cf.~\cite{Kemmer.etal:1959}.
The two-to-one homomorphism from $U(1)\times SU(2)$ to $U(2)$ is given 
by $\bigl(\exp(i\alpha)\,,\,A\bigr)\mapsto\exp(i\alpha)A$ whose kernel 
is $\bigl\{(1,\mathbf{1})\,,\,(-1,-\mathbf{1})\bigr\}$.} 
\begin{subequations}
\label{eq:PauliGroupAction}
\begin{alignat}{3}
\label{eq:PauliGroupAction-a}
\left(\begin{array}{c}
\psi_L\\
\psi^c_R
\end{array}\right)
&\,\mapsto\, \exp\bigl(+i\alpha\bigr)\,
&\left(\begin{array}{cc}
a\ &b\\
-b^*\ &a^*
\end{array}\right)
&\left(\begin{array}{c}
\psi_L\\
\psi^c_R
\end{array}\right)\,,
\\
\label{eq:PauliGroupAction-b}
\left(\begin{array}{c}
\psi^c_L\\
\psi_R
\end{array}\right)
&\,\mapsto\, \exp\bigl(-i\alpha\bigr)\,
&\left(\begin{array}{cc}
a^*\ &b^*\\
-b^{\phantom{*}}\ & a
\end{array}\right)
&\left(\begin{array}{c}
\psi^c_L\\
\psi_R
\end{array}\right)\,.
\end{alignat}
\end{subequations}

Invariance under the Pauli group is now seen to correspond to an 
ambiguity in the particle-antiparticle distinction. This ambiguity 
would only be lifted by interactions that allowed to distinguish 
the two left and the two right states respectively. In the absence 
of such interactions the various definitions of `particle' and 
`antiparticle' are physically indistinguishable, so that the 
Pauli group acts by \emph{gauge} symmetries in the sense of 
Section\,\ref{sec:PhysicalVersusGaugeSymmetries}.
    
Also, the different presentations of the two-component theory, 
already discussed in Section\,\ref{sec:Irritations} can be seen 
here. The Majorana condition reads $\psi=\psi^c$, which in terms 
of the four components introduced above leads to $\psi_L=\psi^c_R$
and $\psi_R=\psi^c_L$. This can be read in two different ways,
depending on whether one addresses $\psi_L,\psi^c_L$ or $\psi_L,\psi_R$
as independent basic states. In the first case one would say that 
there is a left-handed neutrino and its right-handed antiparticle, 
whereas in the second case one regards the tuple $(\psi_L,\psi_R)$
as respectively the left- and right-handed components of a single 
particle which is identical to its antiparticle. 

Beyond weak interaction and beta-decay, the Pauli group played a 
very important r\^ole in Pauli's brief participation in Heisenberg's 
programme for a unified field theory. It was Pauli who first showed  
that the (so far classical) non-linear spinor equation proposed 
by Heisenberg was invariant under the Pauli group 
(cf. Heisenberg's account in his letter to Zimmermann from 
Jan.\,7th 1958 in \cite{Pauli:SC}, Vol.\,IV, Part\,IV\,B, p.\,779).
In this new context the $U(1)$ part of the Pauli group was 
connected to conservation of baryon charge and the $SU(2)$ part
acquired the meaning of isospin symmetry.\footnote{The non-linear 
spinor equation was at that stage not designed to include weak 
interaction.} The central importance of isospin for this programme 
may already be inferred from the title of the proposed common 
publication by Heisenberg and Pauli, which reads: 
\emph{On the Isospin Group in the Theory of Elementary particles}. However, due 
to Pauli's later retreat from this programme, the manuscript 
(cf. \cite{Pauli:SC}, Vol.\,IV, Part\,IV\,B, pp.\,849-861)
for this publication never grew beyond the stage of a preprint.

\subsubsection{Cosmological speculations}
In his last paper on the subject of discrete symmetries, entitled 
\emph{The Violation of Mirror-Symmetries in the Laws of Atomic Physics}\footnote{%
German original: ``Die Verletzung von Spiegelungs-Symmetrien 
in den Gesetzen der Atomphysik''.} (\cite{Pauli:CSP}, Vol.\,2, 
pp.\,1368-1372), Pauli comes back to the question which 
bothered him most: How is the strength of an interaction 
related to its symmetry properties? He says that 
having established a violation of $C$ and $P$ symmetry for weak 
interactions, we may ask why they are maintained for strong and 
electromagnetic interactions, and whether the reason for this 
is to be found in particular properties of these interactions. 
He ends with some speculations on possible connections between 
violations of $C$ and $P$ symmetry in the laws of microphysics
on one hand, and properties of theories of gravitation and its 
cosmological solutions on the other:\footnote{%
German original: ``Zweitens kann man versuchen, einen 
Zusammenhang der Symmetrieverletzungen in Kleinen mit Eigenschaften 
des Universums im Grossen aufzufinden und zu begr\"unden. Dies 
\"uberschreitet aber die M\"oglichkeiten der jetzt bekannten 
Theorien der Gravitation. [...] Um bei der Frage des Zusammenhangs 
zwischen dem Kleinen und dem Grossen \"uber vage Spekulationen 
hinauszugelangen, fehlen daher noch wesentlich neue Ideen. Hiermit 
soll jedoch nicht die Unm\"oglichkeit eines solchen Zusammenhanges
bestimmt behauptet werden.'' (\cite{Pauli:CSP}, Vol.\,2, p.\,1371)}  
\begin{quote}
\emph{
Second, one can try to find and justify a connection between 
symmetry violation in the small with properties of the Universe 
in the large. But this exceeds the capabilities of the 
presently known theory of gravity. [...] New ideas are missing 
to go beyond vague speculations. But this shall not be taken as 
definite expression for the impossibility of such a connection.}
\end{quote}
It may be of interest to contrast this expression of a certain 
open-mindedness for speculations concerning the physics of 
elementary particles on one side and large-scale cosmology 
on the other, with a more critical attitude from Pauli's 
very early writings. In Section\,65 of his Relativity article, 
where Pauli discussed Weyl's attempt for a unifying theory of 
gravity and electromagnetism (to which Pauli himself actively 
contributed), he observes that in Weyl's theory (as well as 
in Einstein's own attempts from that time) it is natural 
to suspect a relation between the size of the electron and 
the size (mean curvature radius) of the universe. But then he 
comments somewhat dismissively that this 
\emph{might seem somewhat fantastic}\footnote{%
German original: ``...was immerhin etwas phantastisch erscheinen 
mag.'' (\cite{Pauli:2000}, p.\,249)}(\cite{Pauli:ToR-Dover}, p.\,202).

\section{Conclusion}
I have tried to display some of the aspects of the notion of 
symmetry in the work of Wolfgang Pauli which to me seem 
sufficiently interesting in their own right. In doing this I 
have drawn freely from Pauli's scientific {\oe}vre, 
irrespectively of whether the particular part is commonly 
regarded as established part of present-day scientific knowledge 
or not. Pauli's faith in the explanatory power of symmetry 
principles clearly shows up in all corner of his {\oe}vre, but 
it also appears clearly rooted beyond the limits of his science.  

In the editorial epilogue to the monumental collection of Pauli's
scientific correspondence, Karl von Meyenn reports that many 
physicists he talked to at the outset of his project spoke against 
the publication of those letters that contained ideas which did not 
stand the test of time~(\cite{Pauli:SC}, Vol.\,IV, Part\,IV\,B, p.\,1375). 
Leaving aside that this must clearly sound outrageous to the historian, 
it is, in my opinion, also totally misguided as far as the scientific 
endeavour is concerned. Science is not only driven by the urge to \emph{know} 
but also, and perhaps most importantly, by the urge to \emph{understand}. 
No one who as ever actively participated in science can deny that. 
One central aspect of scientific understanding, next to offering 
as many as possible alternative and complementary explanations for 
the actual occurrences in Nature, is to comprehend why things could 
not be different from what they appear to be. The insight into a
theoretical or an explanatory failure can be as fruitful as an 
experimental failure. What makes Pauli a great scientist, amongst 
the other most obvious reasons, is not that he did not err---such 
mortals clearly do not exist---, but that we can still learn much 
from where he erred and how he erred. In that sense, let me end by 
the following words from Johann Wolfgang von Goethe's 
\emph{Maximen und Reflexionen} (\#\,1292):

\vspace{0.5cm}
\begin{center}
\doublebox{
\parbox{0.88\linewidth}{
\emph{
\begin{center}
Wenn weise M\"anner nicht irrten, m\"u{\ss}ten die Narren verzweifeln.\\
\medskip
(If wise men did not err, fools should despair.)\\
\medskip
\end{center}
}}}
\end{center}

\newpage
\addcontentsline{toc}{section}{Acknowledgements}
\noindent
\textbf{Acknowledgements:}
I sincerely thank Harald Atmanspacher and Hans Primas for the 
invitation to talk at the conference on \emph{Wolfgang Pauli's Philosophical 
Ideas and Contemporary Science} on the Monte Verita in Ascona, 
Switzerland. I also thank Norbert Straumann for comments and
suggestions for improvement. Finally I wish to express my strongest 
appreciation to the editor of Pauli's Scientific Correspondence, 
Karl von Meyenn, without whose admirable editorial work we would 
not be in the position to share many of Wolfgang Pauli's wonderful 
insights. Thank you!

\addcontentsline{toc}{section}{References}

\end{document}